\documentclass[acmlarge,authorversion]{acmart} 
\usepackage{subfigure,multirow} 
\usepackage[T1]{fontenc}
\usepackage[ruled]{algorithm2e}

\renewcommand{\algorithmcfname}{ALGORITHM}
\SetAlFnt{\small}
\SetAlCapFnt{\small}
\SetAlCapNameFnt{\small}
\SetAlCapHSkip{0pt}
\IncMargin{-\parindent}

\setcopyright{none}

\newcommand{\newtext}[1]{{#1}}
\newcommand{\eg}{e.g.,\ }
\newcommand{\etal}{et al.\ }
\newcommand{\ie}{i.e.,\ }

\begin{document}

\title{Measuring Personalization of Web Search}

\author{ANIK\'{O} HANN\'{A}K}
\affiliation{%
  \institution{Northeastern University}
}
\author{PIOTR SAPIE\.{Z}Y\'{N}SKI}
\affiliation{%
  \institution{Technical University of Denmark}
}
\author{ARASH MOLAVI KAKHKI}
\affiliation{%
  \institution{Northeastern University}
}
\author{DAVID LAZER}
\author{ALAN MISLOVE}
\author{CHRISTO WILSON}
\affiliation{%
  \institution{Northeastern University}
}

\begin{abstract}

Web search is an integral part of our daily lives. Recently, there has been a
trend of personalization in Web search, where different users receive different
results for the same search query. The increasing level of personalization is 
leading to concerns about {\em Filter Bubble} effects, where certain users are simply unable
to access information that the search engines' algorithm decides is irrelevant.
Despite these concerns, there has been little quantification of the extent of
personalization in Web search today, or the user attributes that cause it.

In light of this situation, we make three contributions. First, we develop a
methodology for measuring personalization in Web search results.
While conceptually simple, there are numerous details that our methodology must handle
in order to accurately attribute differences in search results to personalization.
Second, we apply our methodology to 200 users on Google Web Search and 100 users on Bing.
We find that, on average, 11.7\% of results show differences due to personalization on Google,
while 15.8\% of results are personalized on Bing, but that this varies widely by search query
and by result ranking. Third, we investigate the \newtext{user features used to personalize} on Google Web
Search and Bing.  Surprisingly, we only find measurable personalization as a result of 
searching with a logged in account and the IP address of the searching user.
Our results are a first step towards understanding the extent and effects of personalization
on Web search engines today.

\end{abstract}

\terms{Design, Measurement}

\thanks{This article is an extension of the paper ``Measuring Personalization of Web Search'', published at WWW 2013~\cite{hannak-2013-filterbubbles}. This article extends the original by adding measurement results from Bing and DuckDuckGo, as well as adding several new experimental treatments not found in the conference paper.

This research was supported by National Science Foundation grants IIS-0964465, IIS-1408345, and CNS-1054233, and an Amazon Web Services in Education Grant.

Author's addresses: A. Hannak {and} A. Molavi Kakhki {and} D. Lazer {and} A. Mislove {and} C. Wilson, College of Computer and Information Science, Northeastern University; P. Sapie\.{z}y\'{n}ski, Department of Applied Mathematics and Computer Science, Technical University of Denmark.
}

\maketitle

\renewcommand{\shortauthors}{Hann\'{a}k \etal}

\clubpenalty=10000
\widowpenalty=10000

\section{Introduction}
\label{sec:intro}

Web search services like Bing and Google Web Search (Google Search) are an integral part of
our daily lives; Google Search alone receives 17 billion queries per month from U.S.
users~\cite{GoogleSearchStatistics}. People use Web search for a number of reasons,
including finding authoritative sources on a topic, keeping abreast of
breaking news, and making purchasing decisions. The search results that are returned, and their
order, have significant implications: ranking certain results higher or lower can
dramatically affect business outcomes (\eg the popularity of search engine optimization
services), political elections (\eg U.S. Senator Rick Santorum's battle with
Google~\cite{RickSantorumGoogle}), and foreign affairs (\eg Google's ongoing conflict
with Chinese Web censors~\cite{GoogleChina}).

\newtext{Recently, major search engines have implemented {\em personalization}, 
where the Web search operator modifies the results---or their order---based 
on the user who is making the query~\cite{psearch-2005,bing_personal-2011}. 
As previous work has noted~\cite{pretschner-1999-personalize},
an effective personalized search engine is able to decide autonomously whether or
not a user is interested in a specific webpage and, if so, display that result at a higher rank. 
For example, users searching for ``pizza''
in New York and in Boston may receive different results corresponding to local restaurants.
Search engine operators often choose to personalize results as it has been shown 
to provide significant benefits to users (\eg disambiguation of similar
search terms, and retrieval of locally relevant results). 
In fact, the benefits of such personalized rankings has been extensively studied in the research literature 
~\cite{liu-2004-websearch,sieg-2007-ontological,gauch-2003-ontologybased,fan-2000-personalization}.

Unfortunately, while the benefits of personalization are well-studied, the potential negative effects of personalization are not nearly as well-understood.
For example, search engine operators do not typically label which of the returned results were personalized, 
or explain why those results were chosen; only operators themselves know the specifics of 
how the personalization algorithms alter the results. 
Compounding this problem is the fact that measuring personalization in practice poses many challenges, including obtaining large amounts 
of personal data, establishing a baseline for personalization, and distinguishing between inadvertent result changes and personalization.}

As a result, the opaque personalization of Web search has led to growing concerns over the
{\em Filter Bubble} effect~\cite{green-2011-filterbubble}, where users are only given results that the personalization
algorithm thinks they want (while other, potentially important, results remain hidden).
For example, Eli Pariser demonstrated that during the recent Egyptian revolution, different
users searching for ``Tahrir Square'' received either links to news reports of protests, or links
to travel agencies~\cite{pariser-2011-filterbubble}. The Filter Bubble effect is exacerbated by the
dual issues that most users do not know that search results are personalized, yet users
tend to place blind faith in the quality of search results~\cite{pan-2007-trust}.

Concerns about the Filter Bubble effects are now appearing in the popular
press~\cite{singer-2011-filterbubble,sullivan-2012-personalization}, driving
growth in the popularity of alternative search engines that do not personalize results. Unfortunately, to date, there has been little scientific
quantification of the basis and extent of search personalization in practice.

In this paper, we make three contributions towards remedying this situation. {\em First}, we develop a methodology for measuring personalization in Web search results. Measuring personalization is conceptually simple: one can run multiple searches for the same queries and compare the results. However, accurately attributing differences in returned search results to personalization requires accounting for a number of phenomena, including temporal changes in the search index, consistency issues in distributed search indices,  and A/B tests being run by the search provider. We develop a methodology that is able to control for these phenomena and create a command-line-based implementation that we make available to the research community.

{\em Second}, we use this methodology to measure the extent of personalization on multiple popular 
Web search engines: Google Web Search, Bing Search, and DuckGo.\footnote{\newtext{DuckDuckGo is a relatively new search engine that claims to not track users or personalize results. As such, we do not expect to see personalized results, and we include our measurements of DuckDuckGo primarily as a baseline to compare Google Web Search and Bing Search against.}}
We recruit 300 users with active Google and Bing accounts from Amazon's Mechanical Turk to run a list of Web
searches, and we measure the differences in search results that they are given. We control for differences in time,
location, distributed infrastructure, and noise, allowing us to attribute any differences observed to personalization. 
Although our results are only a lower bound, we observe significant personalization:
on average, 11.7\% of Google Web Search results and 15.8\% of Bing Search results show differences
due to personalization, with higher probabilities for results towards the bottom of the page.
We see the highest personalization for queries related to political issues, news, and local
businesses.
We do not observe any noticeable personalization on DuckDuckGo.

{\em Third}, we investigate the \newtext{user features used to personalize}, covering user-provided profile
information, Web browser and operating system choice, search history, search-result-click history, and browsing history.
We create numerous Google and Bing accounts and assign each a set of unique behaviors. We develop a
standard list of 120 search queries that cover a variety of topics pulled from Google Zeitgeist~\cite{GoogleZeitgeist}
and WebMD~\cite{WebMDSearches}. We then measure the differences in results that are returned for this list of
searches. Overall, we find that while the level of personalization is significant, there are very few user properties
that lead to personalization. Contrary to our expectations, for both Google and Bing, we find that only being
logged in to the service and the location (IP address) of the user's machine result in measurable personalization.
All other attributes do not result in level of personalization beyond the baseline noise level.

We view our work as a first step towards measuring and addressing the increasing level of personalization on 
the Web today. All Web search engines periodically introduce new techniques, thus any particular findings about the 
level and \newtext{user features used to personalize} may only be accurate for a small time window. However, our methodology can be 
applied periodically to determine if search services have changed. Additionally, although we focus on Web Search 
in this paper, our methodology naturally generalizes to other search services as well (\eg news, products).  Finally, our methodology may
be useful for communities that routinely use search engines as part of their methodology (e.g., natural language processing or recommendation
systems).

\paragraph{Roadmap} The remainder of this paper is organized as follows: in Section~\ref{sec:background}
we provide a background on Web search and personalization, and a discussion of related work. In Section~\ref{sec:methodology} we describe
our experimental methodology. In Section~\ref{sec:turk} we quantify real-world search personalization
using results from crowdsourced workers, while in Section~\ref{sec:measure} we perform controlled
experiments to ascertain what features search engines use to personalize results. Next, in
Section~\ref{sec:practical}, we examine how the personalization varies over time, across query
categories, and by result rank. We conclude with a discussion of results, limitations, and future
work on Section~\ref{sec:conclusion}.


\section{Background and Related Work}
\label{sec:background}

\newtext{We begin by providing an overview of web search and personalization, followed by a detailed discussion of related work in the research literature.}

\subsection{Background}

Web search services are some of the most popular destinations on the Web; according to Alexa, 
Google and Bing are currently the 2nd and 16th most popular sites on the Internet. Google serves
billions of search results per day~\cite{google-brain-2013}, and controls roughly 67\% of the US
search market, although Bing has risen to control 17\% of the market~\cite{search_share-2013}.
Alternative search engines like DuckDuckGo and Blekko have experienced some success offering
advanced features to power-users, but they have not yet reached mainstream popularity.

\paragraph{Accounts} As the number and scope of the services provided by Microsoft and Google grew,
they both began unifying their account management architecture. Today, Google Accounts are
the single point of login for all Google services (\eg Gmail, YouTube, Google Calendar).
Once a user logs in to one of these services, they are effectively logged in
to all services. A tracking cookie enables all of Google's services to uniquely identify
each logged in user. As of May 2012, Google's privacy policy allows between-service
information sharing across all Google services~\cite{whitten-2012-privacypolicy}.

Similar to Google, Microsoft provides user accounts for its various services (e.g., Windows Live,
Outlook.com, Xbox LIVE, Skype).  These various accounts have been consolidated into a single ``Microsoft account"
(previously called a Windows Live ID); when a user is signed-in to this account, Bing searches are
tracked and saved.

DuckDuckGo explicitly eschews the notion of accounts and users cannot ``log in"; the company claims to not perform
any tracking or profiling of its users.  

\paragraph{Personalization}
Google first introduced ``Personalized Search''
in 2004~\cite{hines-2004-personalsearch}, and merged this product into Google Search
in 2005~\cite{psearch-2005}. In 2009, Google began personalizing search results for
all users, even those without Google accounts~\cite{horling-2009-personalsearch}. 
Recently, Google started
including personalized content from the Google+ social network into search
results~\cite{singhal-2012-personalsearch}.
For example, users may see Web pages which were shared or ``+1'd'' by people in their
Google+ circles alongside normal Google search results.

Bing Search introduced ``Localized Results"~\cite{bing_personal-2011} and 
``Adaptive Search"~\cite{crook-2011-bingadaptive} in 2011, customizing search results using the 
user's location and previous search history, respectively.  In 2013, Bing added ``Social Results"
to the results returned to the user, presenting links shared by a user's Facebook friends alongside
normal search results~\cite{craver-2013-bingsocial}.

DuckDuckGo is explicitly designed to {\em not} personalize results.

\paragraph{Advertising and User Tracking} Both Google and Microsoft are capable of
tracking users as they browse the Web due to their large advertising networks (\eg
DoubleClick and MSN Advertising). Roesner \etal provide an excellent overview
of how Google can use cookies from DoubleClick and Google Analytics, as well as widgets 
from YouTube and Google+ to track users' browsing habits~\cite{roesner-2012-tracking}.  \newtext{Similarly, recent work~\cite{Panopticlick} has shown that it is possible to do {\em browser fingerprinting} (essentially looking for unique browser-agents and other headers) as a method for sites to track returning users.} In theory, data from any of these systems could be used to drive Web search personalization algorithms.

\subsection{Related Work}

Personalized search has been extensively studied in the
literature~\cite{pretschner-1999-personalize,liu-2002-categories,pitkow-2002-personalized,shen-2005-modeling,teevan-2005-personalized,sun-2005-personalized,qiu-2006-personalized,tan-2006-personalized}.
Dou~\etal provide a comprehensive overview of techniques for personalizing
search~\cite{dou-2007-personalized}.
They evaluate many strategies for personalizing search, and conclude that mining user click
histories leads to the most accurate results. In contrast, user profiles have low utility.
The authors also note that personalization is not useful for all types of queries.
\newtext{A recent study investigated the inherent biases of search engines and 
their impact on the quality of information that reaches people~\cite{white-2013-websearchbias} 
They show that the combined effect of people's preferences and the system's inherent bias results 
in settling on incorrect beliefs about half of the time.} 

Other features besides click history have been used to power personalized search. Three
studies leverage geographic location to personalize search~\cite{andrade-2006-relevance,yu-2007-geographic,yi-2009-geointention}. 
Two studies have shown that user demographics can be reliably
inferred from browsing histories, which can be useful for personalizing
content~\cite{hu-2007-predicting,goel-2012-icwsm}. To our knowledge, only one study has
investigated privacy-preserving personalized search~\cite{xu-2007-private}. Given growing
concerns about the Filter Bubble effects, this area seems promising for future research.

Several studies have looked at personalization on systems other than search. Two studies have
examined personalization of targeted ads on the Web~\cite{guha-2010-ads,wills-2012-ads}.
One study examines discriminatory pricing on e-commerce sites, which is essentially
personalization of prices~\cite{mikians-2012-prices}.

In contrast to prior research which focused on building personalized Web services, work is
now emerging that has similar goals to this study, \ie quantifying personalization on
deployed Web services. Majumder \etal propose a latent variable
model to infer the features that a service provider is using to personalize content for
a given user~\cite{majumder-2013-personalization}.

However, there is very little concrete information about how the biggest search engines 
such as Google and Bing personalize their search results.
A 2011 post on the official Google blog states that Google Search personalizes
results based on the user's language, geolocation, history of search queries, and their
Google+ social connections~\cite{singhal-2011-personalsearch}. The specific
uses of search history data are unclear: the blog post suggests that the temporal
order of searches matters, as well as whether users click on results. Similarly, the
specific uses of social data from Google+ are unknown.

Several studies have examined the differences between
results from different search engines. Two studies have performed user studies to compare
search engines~\cite{vaughan-2004-measurements,barilan-2007-rankings}.
Although both studies uncover significant differences between competing search engines,
neither study examines the impact of personalization. Sun \etal propose a method for
visualizing different results from search engines that is based on expected weighted
Hoeffding distance~\cite{sun-2010-visualizing}. Although this
technique is very promising, it does not scale to the size of our experiments.

\begin{figure}
\begin{minipage}{0.49\textwidth}
\centering \includegraphics[width=0.9\textwidth]{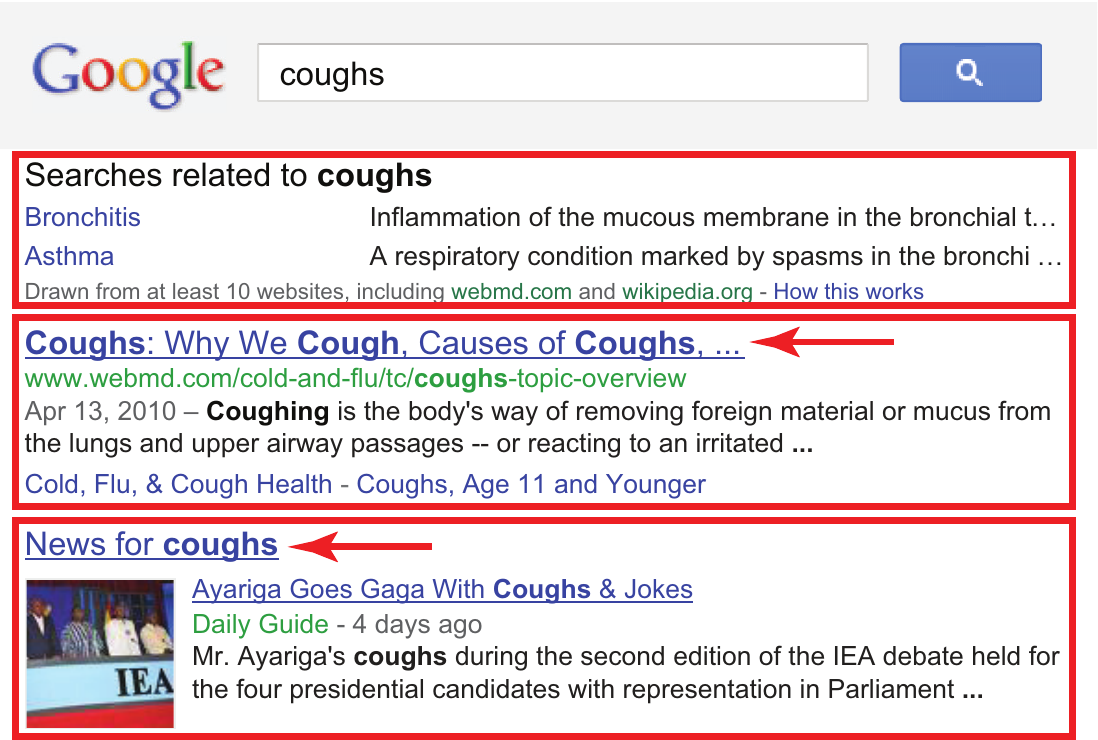}
\caption{Example page of Google Search results.}
\label{fig:google-example}
\end{minipage}
\hfill
\begin{minipage}{0.49\textwidth}
\centering \includegraphics[width=0.9\textwidth]{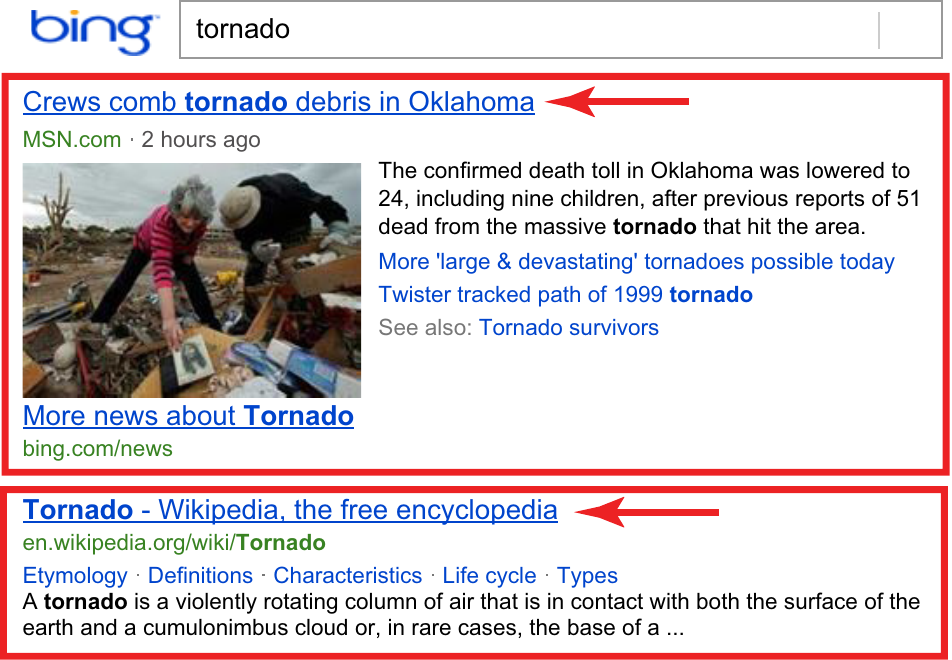}
\caption{Example page of Bing Search results.}
\label{fig:bing-example}
\end{minipage}
\end{figure}


\section{Methods}
\label{sec:methodology}

In this section, we describe our experimental methodology.
First, we give the high-level intuition that guides the design of
our experiments, and identify sources of noise that can lead
to errors in data collection. Second, we describe the implementation
of our experiments. Lastly, we introduce the queries we use to
test for personalization. 

\subsection{Terminology}
\label{sec:terminology}

In this study, we use a specific set of terms when referring to Web search. 
Each {\em query} to a Web search engine is composed of one or more keywords. In
response to a query, the search engine returns a {\em page} of {\em results}.
Figure~\ref{fig:google-example} shows a truncated example page of Google Search
results for the query ``coughs'', and Figure~\ref{fig:bing-example} shows a truncated example
page of Bing Search results for the query ``tornado."  Each page contains $\approx$10 results
(in some cases there may be more or less). We highlight three results with red boxes in both
figures. Most results contain $\ge1$ links. In this study, we only focus on the {\em primary link} in
each result, which we highlight with red arrows in Figures~\ref{fig:google-example} and~\ref{fig:bing-example}.

In most cases, the primary link is {\em organic}, \ie it points to a third-party
website ~\cite{chan-2012-organic}. The WebMD result in Figure~\ref{fig:google-example}
falls into this category. However, the primary link may point to another Google or Microsoft
service. For example, in Figure~\ref{fig:google-example}
the ``News for coughs'' link directs to Google News, and the ``More news about Tornado" link
in Figure~\ref{fig:bing-example} directs to Bing News. \newtext{Search engines often
include links to other {\em services} offered by the same company; this strategy is
sometimes referred to as ``aggregated search.''}

A few services inserted in Web search results do not include a
primary link. The ``Related Searches'' result in Figure~\ref{fig:google-example}
falls into this category. Another example is Google Dictionary, which displays
the definition of a search keyword. In these cases, we treat the primary link
of the result as a descriptive, static string, \eg ``Related'' or ``Dictionary.''

\paragraph{DuckDuckGo} Search results from DuckDuckGo follow a different format from
Google and Bing. On DuckDuckGo, the top of the search result page is dominated by a box
of contextual information related to the query. For example, after searching for
``barack obama'' the contextual box contains information about the president taken
from Wikipedia, and links to recent news articles. Below the contextual box is the
list of organic search results. Unlike Google and Bing, DuckDuckGo does not return
multiple different pages of search results. Instead, the page continually loads more results
as the user scrolls down the page.

In this study, we focus on the search results returned by DuckDuckGo, and ignore links
in the contextual box. On DuckDuckGo, \newtext{results are presented in a simple ordered list,
so there is no problem of having multiple links in one result.} We focus 
on the top 10 results returned by DuckDuckGo, so that the analysis is comparable
across the three search engines.

\subsection{Experiment Design}
\label{sec:design}

Our study seeks to answer two broad questions. First,
{\em what user features influence Web search personalization algorithms?}
This question is fundamental: outside of Web search companies, nobody knows the
specifics of how personalization works.
Second, {\em to what extent does search personalization actually affect
search results?} Although it is known that Web search companies personalize search
results, it is not clear how much these algorithms actually alter the
results. If the delta between ``normal'' and ``personalized'' results
is small, then concerns over the Filter Bubble effect may be misguided.


At a high-level, our methodology is to execute carefully controlled
queries on different Web search engines to identify what user features
trigger personalization. Each experiment follows a similar pattern: first,
create $x$ accounts that each vary
by one specific feature. Second, execute $q$ identical queries from
each account, once per day for $d$ days. Save the results of each
query. Finally, compare the results of the queries to
determine whether the same results are being served in the same order to
each account. If the results vary between accounts,
then the changes can be attributed to personalization linked to the
given experimental feature. Note that certain experimental treatments are run
{\em without} accounts (\ie to simulate users without accounts). Furthermore,
some Web search providers do not allow users to create accounts (\eg DuckDuckGo).

\paragraph{Sources of Noise} Despite the simplicity of the high-level
experimental design, there are several sources of noise that can
cause identical queries to return different results.

\begin{itemize}
\item {\bf Updates to the Search Index:} Web search services constantly update their
search indices. This means that the results for a query may
change over time.

\item {\bf Distributed Infrastructure:} Large-scale Web search services
are spread across geographically diverse datacenters. Our tests have shown that
different datacenters may return different results for the same
queries. It is likely that these differences arise due to
inconsistencies in the search index across datacenters.

\item {\bf Geolocation:} Search engines use the user's IP address
to provide localized results~\cite{yi-2009-geointention}.  Thus, searches
from different subnets may receive different results.

\item {\bf A/B Testing:} Web search services sometimes conduct A/B 
testing~\cite{pansari-2006-abtest}, where certain results are altered
to measure whether users click on them more often. Thus, there may be a
certain level of noise independent of all other factors.

\end{itemize}

\paragraph{The Carry-Over Effect} One particular source of \newtext{noise comes from
the dependency of searches within one "browsing session."
For example, if a user searches for query $A$, and then searches for query $B$,
the results for $B$ may be influenced by the previous search for $A$.
Prior research on user intent while searching has shown that sequential queries
from a user are useful for refining search result
~\cite{cheng-2010-intent,shen-2011-intent,mobasher-2000-webusage,andersen-2000-clickstreams,smith-2005-content}.
Thus, it is not surprising that some search engines implement {\em query refinement}
using consecutive keywords within a user's browsing session.
We term the effect of query refinement on subsequent searches as the {\em carry-over effect}.}

\begin{figure*}[t]
\begin{minipage}{0.48\textwidth}
  \centering \includegraphics[width=1.0\textwidth]{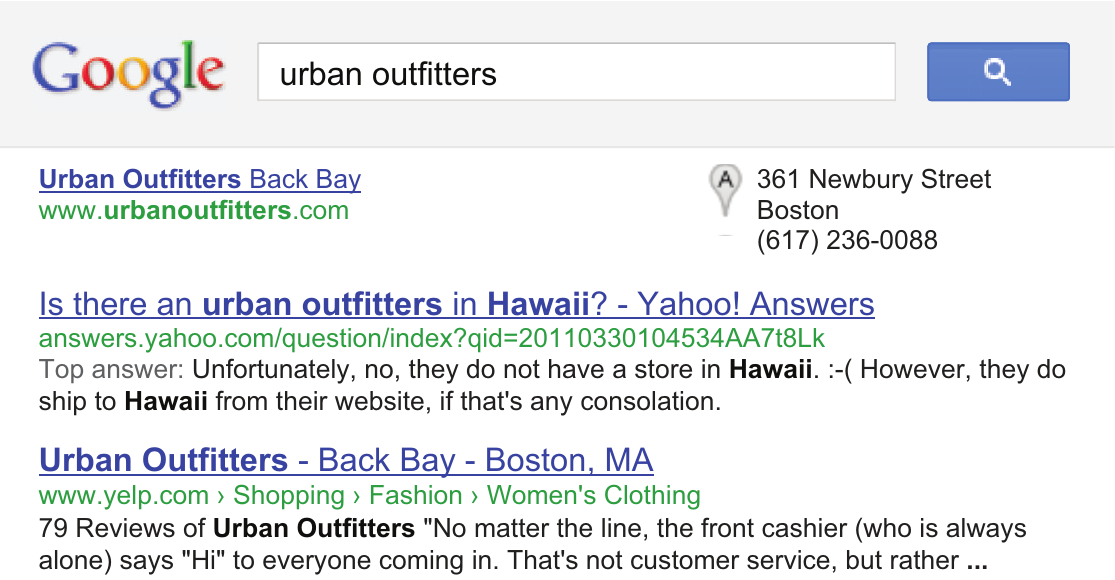}
  \caption{Example of result carry-over, searching for ``hawaii" then searching for ``urban outfitters."}
  \label{fig:carry-over}
\end{minipage}
\hfill
\begin{minipage}{0.48\textwidth}
  \centering \includegraphics[width=1.0\textwidth]{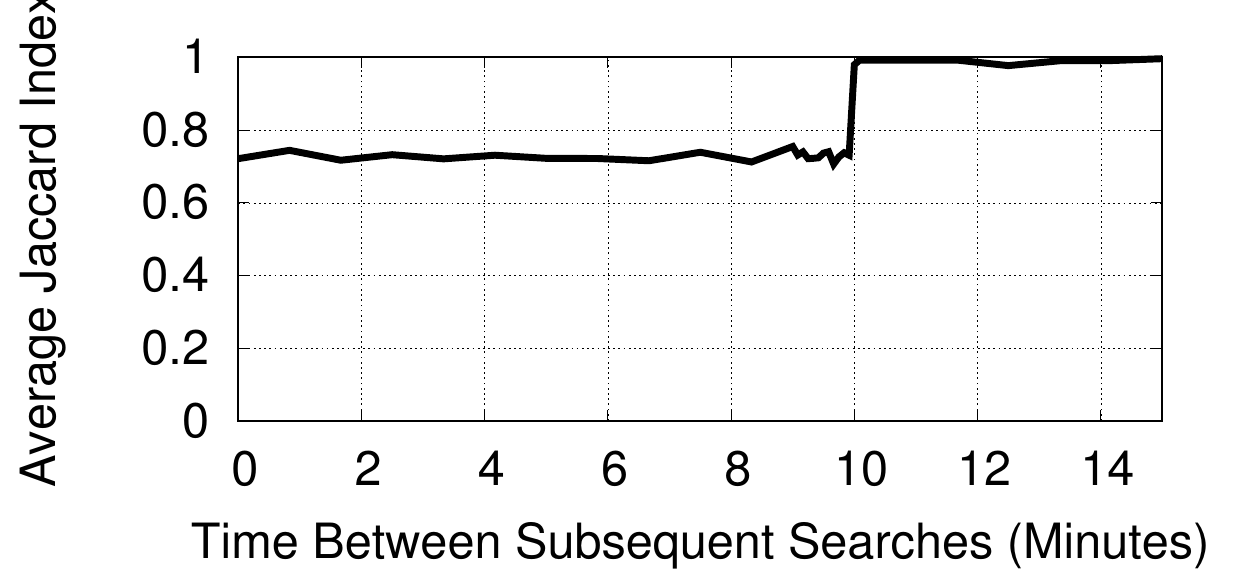}
  \caption{Overlap of results when searching for ``test" followed by ``touring''
  compared to just ``touring'' for different waiting periods.}
  \label{fig:carryover-exp}
\end{minipage}
\end{figure*}

An example of carry-over on Google Search is shown in Figure~\ref{fig:carry-over}. In
this test, we search for ``hawaii'' and then immediately search
for ``urban outfitters'' (a clothing retailer). We conducted
the searches from a Boston IP address, so the results include links
to the Urban Outfitters store in Boston. However, because
the previous query was ``hawaii,'' results pertaining to Urban Outfitters
in Hawai'i are also shown.

To determine how close in time search queries must be to trigger carry-over,
we conduct a simple experiment. We first pick different pairs of queries
(\eg ``gay marriage" and ``obama").  We then start two different browser
instances: in one we search for the first query, wait, and then for the second query,
while in the other we search only for the second query.  We repeat this experiment
with different wait times, and re-run the experiment 50 times with different query
pairs. Finally, we compare the results returned in the two different browser instances 
for the second term.

The results of this experiment on Google Search are shown in
Figure~\ref{fig:carryover-exp} for the terms ``test" and ``touring" (other pairs of
queries show similar results). The carry-over effect can be clearly observed: the
results share, on average, seven common results (out of 10) when the interval between
the searches is less than 10 minutes (in this case, results
pertaining to Turing Tests are included).  After 10 minutes, the carry-over effect
disappears. Thus, in all Google-focused experiments in the following sections, we wait at
least 11 minutes between subsequent searches in order to avoid any carry-over effects.
In our testing, we observed carry-over for both logged in users and users without
Google accounts.

We performed the same experiments on Bing and DuckDuckGo, but did not observe any
carry-over effects. Thus, we conclude that the carry-over effect is unique to Google Search
(at least in fall 2012, when we were conducting measurements).

\paragraph{Controlling Against Noise} In order to mitigate measurements
errors due to these factors, we perform a number of steps (some borrowed from~\cite{guha-2010-ads}):
{\em First}, all of our queries are executed by the normal Web search
webpage, rather than via any search APIs. It has been shown that search engine
APIs sometimes return different results than the standard webpage~\cite{crown-2007-disagree}.
{\em Second}, all of our machines execute
searches for the same query at the same time (\ie in lock-step).
This eliminates differences in query results due to temporal effects.
This also means that each of our accounts has exactly the same
search history at the same time.
{\em Third}, we use static DNS
entries to direct all of our query traffic to a specific Web search provider IP
address. This eliminates errors arising from differences between
datacenters.
{\em Fourth}, we wait 11 minutes in-between subsequent queries to avoid carry-over.
As shown in Figure~\ref{fig:carryover-exp}, an 11 minute wait is sufficient to avoid
the majority of instances of carry-over. For consistency, we use this same methodology
for Google Search, Bing, and DuckDuckGo, even though the latter two do not exhibit carry-over.
{\em Fifth}, unless otherwise stated, we send all of the search queries for a given
experiment from the same /24 subnet. Doing so ensures that any geolocation would
affect all results equally.

{\em Sixth}, we include a
{\em control account} in each of our experiments. The control account
is configured in an identical manner to one other account in the given
experiment (essentially, we run one of the experimental treatments twice). 
By comparing the results received by the control and its
duplicate, we can determine the baseline level of noise in the
experiment (\eg noise caused by A/B testing). Intuitively, the control
should receive exactly the same
search results as its duplicate because they are configured identically,
and perform the same actions at the same time. If there is divergence
between their results, it must be due to noise. 

\subsection{Implementation}


Our experiments are implemented using
custom scripts for PhantomJS~\cite{PhantomJS}. We chose
PhantomJS because it is a full implementation of the WebKit browser,
\ie it executes JavaScript, manages cookies, {\em etc.}
Thus, using PhantomJS is significantly more realistic than using custom code that does not
execute JavaScript, and it is more scalable than automating a full
Web browser (\eg Selenium~\cite{Selenium}).

On start, each PhantomJS instance logs in to a Web search account (\eg a Google or Microsoft account)
using separate credentials, and begins issuing queries to the Web search engine. The script downloads
the first page of search results for each query. The script waits 11 minutes
in-between searches for subsequent queries.

During execution, each PhantomJS instance remains persistent in memory and
stores all received cookies. After executing all assigned queries,
each PhantomJS instance closes and its cookies are cleared. The various cookies
are recreated during the next invocation of the experiment when the script
logs in to its assigned account. All of our experiments are designed
to complete in $\approx$24 hours.

All instances of PhantomJS are run on a single machine. We modified the
{\tt /etc/hosts} file of this machine so that DNS queries to Web search services resolve
to specific IP addresses. We use SSH tunnels to forward traffic from
each PhantomJS instance to a unique IP address in the same /24 subnet.


All of our experiments were conducted in fall of 2012 and spring of 2013. Although our results
are representative for this time period, they may not hold in the
future, since Web search engines are constantly tweaking their personalization algorithms.

\paragraph{Accounts} Unless otherwise specified, each Google and Microsoft account we
create has the same profile: 27 year old, female. The default User-Agent
is Chrome 22 on Windows 7. As shown in Section~\ref{sec:basics},
we do not observe any personalization of results based on these attributes.

We manually crafted each of our accounts to minimize the likelihood
of being automatically detected. Each account was given a unique
name and profile image. We read all of the introductory emails in each
account's email inbox (\ie in GMail or Hotmail). To
the best of our knowledge, none of our accounts were banned or flagged by Google or Microsoft
during our experiments.

\begin{table}
\caption{Categories of search queries used in our experiments.}
\label{tab:queries}
\begin{tabular}{l|l|l}
{\bf Category} & {\bf Examples} & {\bf No.}\\
\hline
Tech & Gadgets, Home Appliances & 20 \\
News & Politics, News Sources & 20 \\
Lifestyle & Apparel Brands, Travel Destinations, Home and Garden & 30\\
Quirky & Weird Environmental, What-Is? & 20 \\ 
Humanities & Literature & 10 \\
Science & Health, Environment & 20 \\
\hline
{\bf Total} & & 120 \\
\end{tabular}
\end{table}

\subsection{Search Queries}

In our experiments, each account searches for a specific
list of queries. It is fundamental to our research that we select a list
of queries that has both breadth and impact. Breadth is vital, since
we do not know which queries Web search engines personalize results for. However,
given that we cannot test all possible queries, it is important that we
select queries that real people are likely to use.

\newtext{Traditionally, search queries are classified into three different classes according to
their intent: navigational, informational and transactional~\cite{broder-2002-taxonomy}.
Navigational queries are not interesting from the perspective of personalization, since
navigational queries tend to have a single, ``correct'' answer, \ie the URL of the desired
website. In contrast, the results of informational and transactional queries could be personalized;
in both cases, the user's intent is to seek out information or products from a potentially
large number of websites. Thus, in our experiments we focus on informational and transactional
queries. }

As shown in Table~\ref{tab:queries}, we use 120 queries divided equally over
12 categories in our experiments. These queries were chosen
from the 2011 Google Zeitgeist~\cite{GoogleZeitgeist}, and WebMD~\cite{WebMDSearches}.
Google Zeitgeist is published annually by Google, and highlights the most
popular search queries from the previous calendar year. We chose these queries
for two reasons: first, they cover a broad range of categories (breadth). Second,
these queries are popular by definition, \ie they
are guaranteed to impact a large number of people.

The queries from Google Zeitgeist cover many important
areas. 10 queries are political (\eg ``Obama Jobs Plan'', ``2012 Republican Candidates'')
and 10 are related to news sources (\eg ``USA Today News'').
Personalization of political and news-related searches are some of the most 
contentious issues raised in Eli Pariser's book on the Filter Bubble effects~\cite{pariser-2011-filterbubble}.
Furthermore, several categories are shopping related (\eg gadgets,
apparel brands, travel destination). As demonstrated by Orbitz, shopping related
searches are prime targets for personalization~\cite{mattioli-2012-orbitz}.

One critical area that is not covered by Google Zeitgeist is health-related
queries. To fill this gap, we chose ten random queries from
WebMD's list of popular health topics~\cite{WebMDSearches}.

\section{Real-World Personalization}
\label{sec:turk}


We begin by measuring the extent of personalization that users are seeing today. Doing so
requires obtaining access to the search results observed by real users; we therefore
conducted a simple user study.

\subsection{Collecting Real-World Data}

We posted two tasks on Amazon's Mechanical Turk (AMT), explaining our study and
offering each user \$2.00 to participate.\footnote{This study was conducted under Northeastern University IRB protocol \#12-08-42; all personally identifiable information was removed from the dataset.} In the first task, participants were required to 1) be in the United States, 2) have a Google account, and 3) be logged in to Google during the study. The second task was analogous to the first, except it targeted users with Bing accounts. Users who accepted either task were instructed to configure their Web browser to use a HTTP proxy controlled by us. \newtext{Then, the users were directed to visit a Web page hosted on our research server.  This page contained JavaScript} that automatically performed \newtext{the same 80 searches} on Google or Bing, respectively.\footnote{We make the source code for this page available to the research community so that our experiment can easily be replicated.} 50 of the queries were randomly chosen from the categories in Table~\ref{tab:queries}, while 30 were chosen by us. 

The HTTP proxy serves several functions. {\em First}, the proxy records the search engines' HTML responses to the users' queries so that we can observe the results returned to the user. \newtext{We refer to these results as the {\em experimental results}.}  {\em Second}, each time the proxy observes a user making a query, it executes two PhantomJS scripts. Each script logs in to the respective search engine and executes the same exact query as the user. \newtext{We refer to the results observed by these two scripts as the {\em control results}, and} they allow us to compare results from a real user (who Google/Bing has collected extensive data on) to fresh accounts (that have minimal Google/Bing history). {\em Third}, the proxy controls for noise in two ways: 1) by executing user queries and the corresponding scripted queries in parallel, and 2) forwarding all search engine traffic to hard-coded IP addresses for Google and Bing.

\begin{figure*}[t!]
       \includegraphics[width=\textwidth]{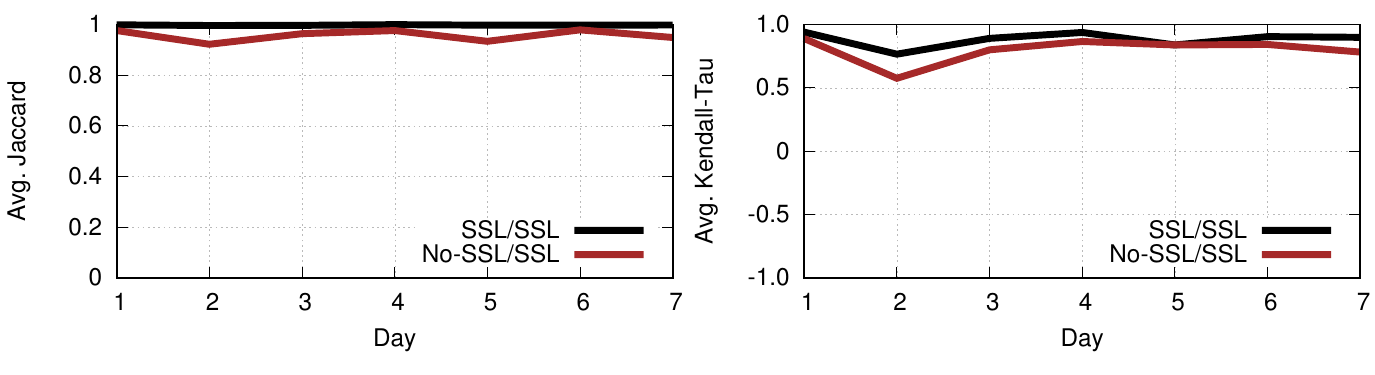}
        \caption{Results for the no-SSL versus SSL experiment on Google Search.}
        \label{fig:ssl_exp}
\end{figure*}

\paragraph{SSL versus no-SSL} Although the proxy is necessary to control for noise, there is a caveat to this technique
when it is applied to Google Search. Queries from AMT users must be sent to {\tt http://google.com}, whereas the controls use {\tt https://google.com}. The reason for this issue is that  HTTPS Google Search rejects requests from proxies, since they could indicate a man-in-the-middle attack. Unfortunately, result pages from HTTP Google Search include a disclaimer explaining that some types of search personalization are disabled for HTTP results.

\newtext{To understand if the differences between SSL and no-SSL Google Search are significant, we conducted a simple pilot study. We automated three Firefox browsers to execute our 120 search queries every day for seven days. Two of the browsers searched using {\tt https://google.com}, and the third searched on {\tt http://google.com} (\ie SSL search serves as the control for this experiment). The three browsers were sandboxed so they could not influence each other (\eg via cached files or cookies), and all cookies and history were cleared from the browsers before beginning the experiment. 

Figure~\ref{fig:ssl_exp} shows the average Jaccard Index and average Kendall's Tau for each day of test results. Both quantities are averaged over all 120 queries. The ``SSL/SSL'' line compares the results received by the two accounts that searched using {\tt https://google.com}. As expected, the results received by the accounts have the same composition (\ie Jaccard Index is 0.998 on average), although the order of results is somewhat noisy (\ie Kendall's Tau is 0.88 on average). The ``No-SSL/SSL'' line compares the results received by the account that searched using {\tt http://google.com} to an account that searched using {\tt https://google.com}. The results show that there are consistent, but minor, differences between the composition and ordering of the two search results. Average Jaccard and Kendall't Tau are 0.95 and 0.79 for the the no-SSL/SSL experiments, respectively.

The takeaway from Figure~\ref{fig:ssl_exp} is that there are slight differences in the search results from SSL and no-SSL Google Search. However, the variation induced by noise is greater than the variation induced by the presence or absence of encryption. Thus, we feel that the experimental methodology used in this section is sound overall, because we are able to control for changes in search results due to noise.

\paragraph{Alternate Methodologies} Other researchers have developed alternate techniques to compare search results across users. For example, the authors of the ``Bing it On'' study~\cite{BingItOn} had users take screenshots of search results and uploading them to the experimenters.  We found such an approach to be  a poor fit for our experimental goals, as requesting users to submit screenshots for every search would (a) significantly reduce the coverage of search terms (since users would have to manually upload screenshots, instead of the searches being automatic) and (b) make it more difficult to control for noise (since it would not be possible to run the user query and the control query in lock-step).
}

\begin{figure*}[t]
  \includegraphics[width=1.0\textwidth]{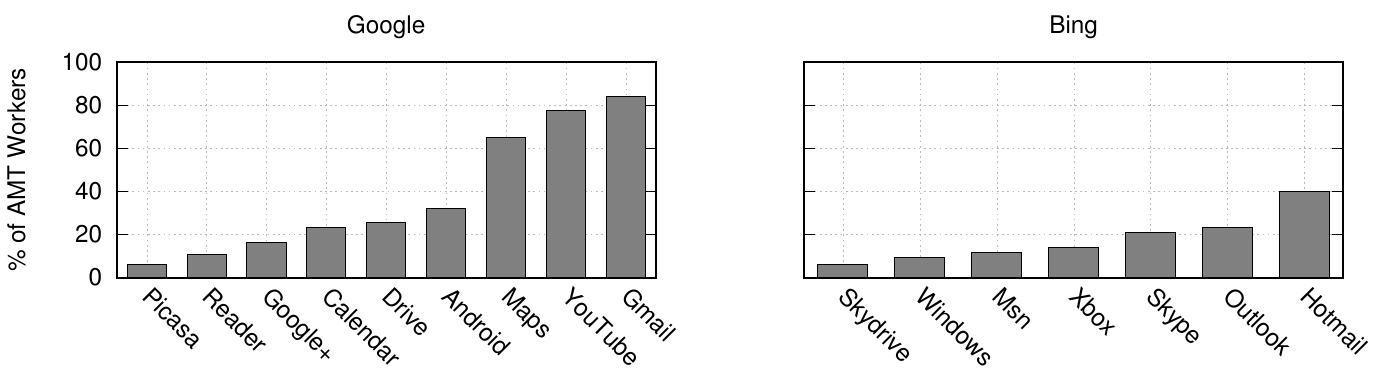}
  \caption{Usage of Google/Microsoft services by AMT workers.}
  \label{fig:survey}
\end{figure*}

\begin{figure*}[t]
  \includegraphics[width=1.0\textwidth]{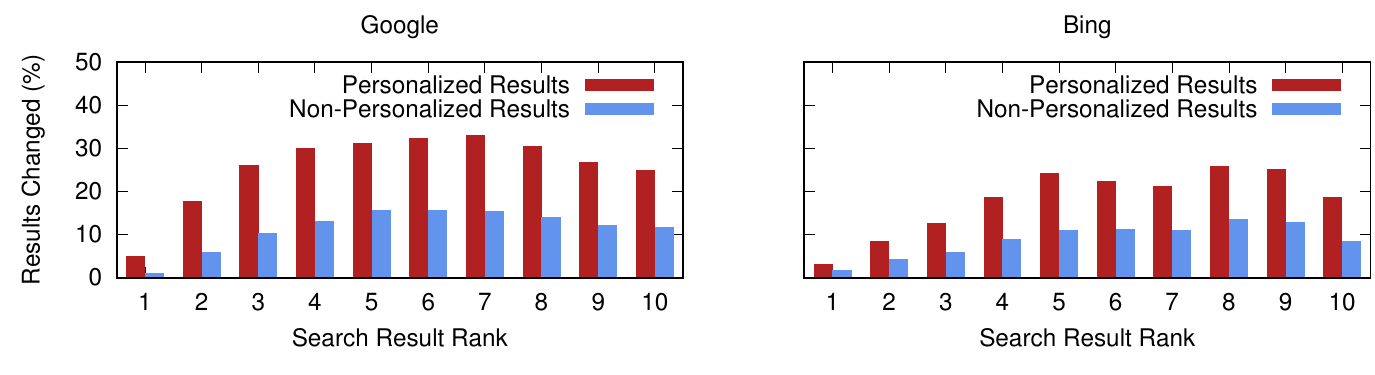}
  \caption{\% of AMT and control results changed at each rank.}
  \label{fig:turk_rank}
\end{figure*}

\paragraph{AMT Worker Demographics} In total, we recruited 300 AMT workers, 200 for our 
Google Search experiment and 100 for our Bing experiment.  \newtext{The reason for fewer users in the Bing experiment is
that we were only able recruit 100 AMT workers who hold Bing accounts (it appears that Bing accounts are much less common).} 
In both experiments, the participants first answered a brief demographic survey.
Our participants self-reported to residing in 43 different U.S. states, and range in age from 
18 to 66 (with a bias towards younger users). Figure~\ref{fig:survey} shows the usage of Google
and Microsoft services by our participants. For Google, 84\% are Gmail users, followed by 76\%
that use YouTube, while for Bing 40\% are Hotmail users. These survey results 
demonstrate that our participants 1) come from a broad sample of the U.S. population, and 2)
use a wide variety of Google and Microsoft services. The low usage of Microsoft
Windows may be due to issues experienced by Internet Explorer users: written feedback
from several of our participants indicated that Internet Explorer users found it difficult to
set up the necessary proxy settings for our tasks.

  \begin{table}[b]
  \caption{Top 10 most/least personalized queries on Google Search and Bing.}
  \label{tab:top10turks}
    \begin{tabular}{r l l | l l}
    \multicolumn{3}{c|}{\bf Most Personalized} & \multicolumn{2}{|c}{\bf Least Personalized} \\
    \hline
    & {\bf Google} & {\bf Bing} & {\bf Google} & {\bf Bing} \\
    \hline
    1. & gap & harry & what is gout & what is vegan \\
    2. & hollister & 2008 crysis & dance with dragons & theadvocate \\
    3. & hgtv & nuclear weapon & what is lupus & arash molavi \\
    4. & boomerang & witch & gila monster facts & hollister \\
    5. & home depot & job creation & what is gluten & osteoporosis \\
    6. & greece & tax cuts & ipad 2 & what is gluten \\
    7. & pottery barn & issue & cheri daniels & hot to dispose of paint \\
    8. & human rights & abortion & psoriatic arthritis & wild kratts \\
    9. & h2o & iran and isreal & keurig coffee maker & gap \\
    10. & nike & obama & maytag refrigerator & amana refrigerator \\
    \end{tabular}
  \end{table}
  
\subsection{Results}

We now pose the question: {\em how often do real users receive personalized search results?}
To answer this question, we compare the results received by AMT users and the corresponding
control accounts. Figure~\ref{fig:turk_rank} shows the percentage of results
that differ at each rank (\ie result 1, result 2, {\em etc.}) when we compare the AMT results
to the control results, and the control results to each other. Intuitively, the percent change
between the controls is the noise floor; any change above the noise floor when comparing AMT
results to the control can be attributed to personalization.

There are three takeaways from Figure~\ref{fig:turk_rank}. First, we observe extensive
personalization of search results. On average, across all ranks, AMT results showed an
11.7\% {\em higher} likelihood of differing from the control result than the controls
results did from each other on Google Search, and 15.8\% higher likelihood on Bing.
This additional difference can be attributed to personalization.
\newtext{To make sure these differences between the AMT and the control results are in fact statistically 
significant (and not just a reflection of the sampling), we perform the Chi squared test. 
We calculate the $p$-value for each rank for both Bing and Google; we find all of the $p$-values to be
lower than 0.0001, indicating the the results are statistically significant.  }
Second, as already indicated, we observe more personalization on Bing than on Google
Search. Third and finally, top ranks tend to be less personalized than bottom ranks on
both search engines.

To better understand how personalization varies across queries, we list the
top 10 most and least personalized queries on Google Search and Bing in Table~\ref{tab:top10turks}.
The level of personalization per query is calculated as the probability of AMT results
equaling the control results, minus the probability of the control results equaling each other.
Large values for this quantity indicate large divergence between AMT and control results, as
well as low noise (\ie low control/control divergence).



As shown in Table~\ref{tab:top10turks}, the most personalized queries on Bing tend to be
related to important political issues (\eg ``job creation'' and ``tax cuts'') whereas on
Google the most  personalized queries tend to be related to companies and politics
(\eg ``greece'', ``human rights,'' and ``home depot''). In contrast, the least personalized
results on both search engines are often factual (``what is'') and health related queries.

We manually examined the most personalized results and observed that most of the personalization
on Google is based
on location. Even though all of the AMT users' requests went through our proxy and thus appeared
to Google as being from the same IP address, Google Search returned results that are specific
to other locations. This was especially common for company names, where AMT users
received results for different store locations.

\section{Personalization Features}
\label{sec:measure}

In the previous section, we observed significant personalization for real users on Google Search and Bing.  \newtext{We would now like to explore which user {\em features} (\ie aspect of the users' profile or activity) are most likely to lead to personalized results.  To do so, we are unable to use existing real user accounts as we did before, as the history of profile attributes and activity of these accounts are unknown to us.  Instead, we create new, synthetic accounts under our control, and use these accounts (whose entire history we do know) to determine which features are most influential.}

Although we cannot possibly enumerate and test all possible user features, we can investigate likely candidates. \newtext{To do so, we enumerated the list of user features that (a) have been suggested in the literature as good candidates for personalization and (b) are possible to emulate given the constraints of our experimental methodology; we discuss the user features we were not able to explore in Section~\ref{sec:conclusion}.} Table~\ref{tab:features} lists the different user features that our experiments emulate, as well as which search engines each user feature was evaluated on.

\subsection{\newtext{Collecting Synthetic Account Data}}

For each user feature we wish to examine, we create $x+1$ fresh user accounts, where $x$ equals the number of possible values of the feature we are testing in that experiment, plus one
additional {\em control account}. We refer to all non-control accounts as {\em test accounts} For example, in the Gender experiment, we create \newtext{four accounts in total: three test accounts (one ``male,'' one ``female,'' one ``other'') and one control account (``female").} We
execute $x+1$ instances of our PhantomJS script for each experiment \newtext{(one for each account)}, and forward
the traffic to $x+1$ unique endpoints via SSH tunnels. Each account searches
for all 120 of our queries, and we repeat this process daily
for 30 days. This complete treatment is conducted on Google, Bing, and DuckDuckGo (depending on the particular feature under analysis).  \newtext{As before, we compare the differences in the results between the control account and its counterpart (in our example above, the two ``female" accounts) to measure the baseline noise; we then compare the differences in the results between the test accounts and the control to measure personalization.}

\newtext{It is important to note that we can not compare results across search engines 
given that their coverage on different topics might vary; thus, our measurements 
aim for capturing the personalization level within each search engine.}


\subsection{Measuring Personalization}
When comparing the list of search results for test and control accounts, we use two
metrics to measure personalization. First, we use Jaccard Index, which views the result lists
as sets and is defined as the size of the intersection over the size of the union. A Jaccard
Index of 0 represents no overlap between the lists, while 1 indicates they contain the same
results (although not necessarily in the same order).  

 \newtext{To measure reordering, we use Kendall's tau rank correlation coefficient. This metric is commonly used
in the information retrieval literature to measure the similarity of the orderings of the data when ranked by each of the quantities.
To calculate Kendall's tau coefficient on two ranked lists we take the difference between 
the number of concordant pairs and the number of 
discordant pairs and normalize it with the number of possible pairings of the two lists. 
If the agreement between two rankings is perfect, the coefficient value is 1.}

\begin{table}
\caption{User features evaluated for effects on search personalization.
The ``Tested On'' column details whether we evaluated the given feature on
Google ({\em G}), Bing ({\em B}), and/or DuckDuckGo ({\em D}).}
\label{tab:features}
\begin{tabular}{l|r|c|l}
{\bf Category} & {\bf Feature} & {\bf Tested On} & {\bf Tested Values} \\
\hline
Tracking & Cookies & G, B & Logged In, Logged Out, No Cookies \\
\multirow{2}{36mm}{User-Agent} & OS & G, B, D & Win. XP, Win. 7, OS X, Linux \\
& Browser & G, B, D& Chrome 22, Firefox 15, IE 6, IE 8, Safari 5 \\
\hline
Geolocation & IP Address & G, B, D & MA, PA, IL, WA, CA, UT, NC, NY, OR, GA \\
\hline
\multirow{3}{36mm}{User Profile} & Gender & G, B & Male, Female, Other \\
& Age & G, B & 15, 25, 35, 45, 55, 65 \\
& Zip Code & B & MA, CA, FL, TX, WA \\
\hline
\multirow{5}{36mm}{Search History,\\Click History,\\and Browsing History} & Gender & G, B & Male, Female \\
& Age &  G, B & $<$18, 18-24, 25-34, 35-44, 45-54, 55-64, $\ge$65 \\
& Income & G, B & \$0-50K, \$50-100K, \$100-150K, $>$\$150K \\
& Education & G, B & No College, College, Grad School \\
& Ethnicity & G, B & Caucasian, African American, Asian, Hispanic\\
\end{tabular}
\end{table}

\subsection{Basic Features}
\label{sec:basics}

We begin our experiments by focusing on features associated with a
user's browser, their physical location, and their user profile. 

\begin{figure*}[t!]
        \includegraphics[width=\textwidth]{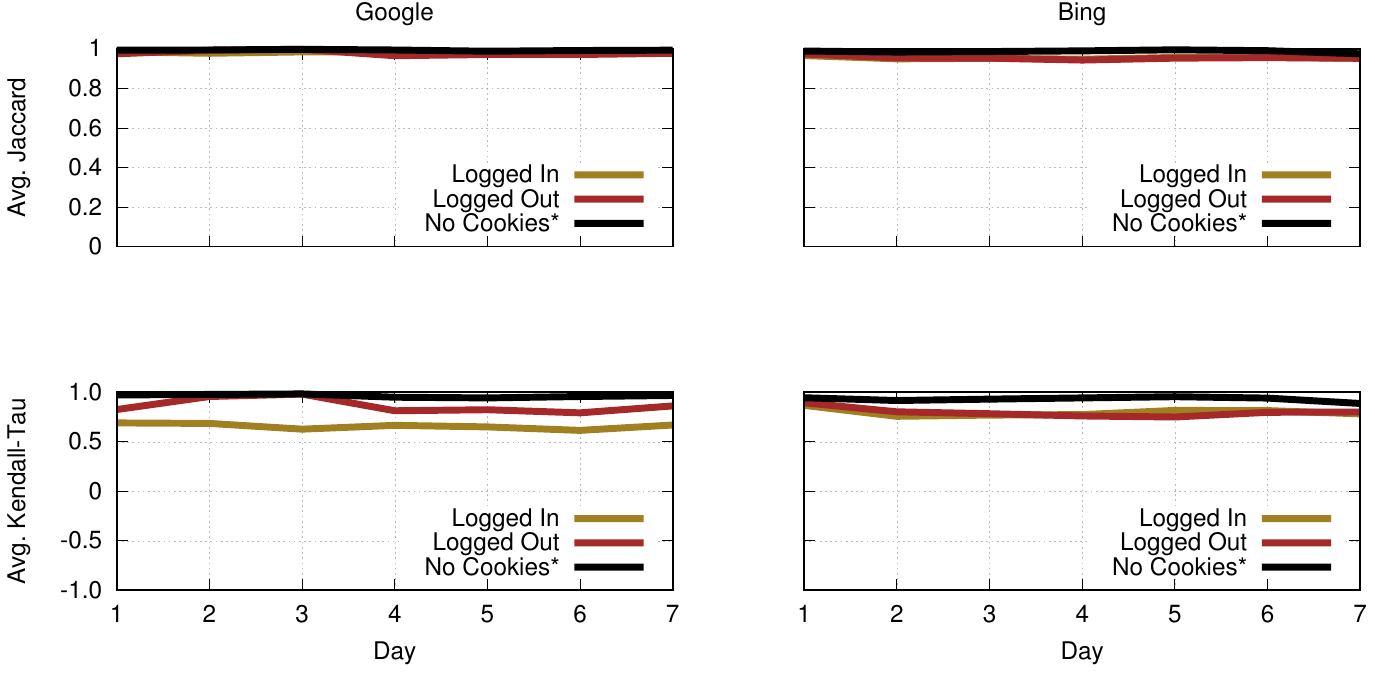}
        \caption{Results for the cookie tracking experiments on Google and Bing.}
        \label{fig:cookie}
\end{figure*}


\paragraph{Basic Cookie Tracking} In this experiment, the goal is to compare the
search results for users who are logged in to a Google/Bing account, not logged in,
and who do not support cookies at all. Google and Bing are able to
track the logged in and logged out users, since both search engines place tracking
cookies on all users, even if they do not have a user account. The user who does
not support cookies receives a new tracking cookie after every request, and
we confirm that the identifiers in these cookies are unique on
every request. However, it is unknown whether Google or Bing are able to link these new
identifiers together behind-the-scenes (\eg by using the user's IP address as
a unique identifier).

To conduct this experiment, we use four instances of PhantomJS per search engine.
The first two completely clear their cookies after every request. The third
account logs in to Google/Bing and persists cookies normally. The fourth account does
not log in to Google/Bing, and also persists cookies normally.

Figure~\ref{fig:cookie} shows the results of our experiments. The upper left plot
shows the average Jaccard Index for each
account type (logged in/logged out/no cookies) across all search queries on Google
when compared to the control (no cookies).
In all of our figures, we place a {\em *} on the legend entry that corresponds to the control
test, \ie two accounts that have identical features. The figure reveals that the results
received by users are not dependent on whether they support cookies, or their login state
with Google. However, just because the results are the same, does not mean that they
are returned in the same order.

To examine how the order of results changes, we plot the average Kendall's tau coefficient between
each account type versus the control on Google in the lower left plot of
Figure~\ref{fig:cookie}. We observe that a user's login state and
cookies do impact the order of results from Google Search. The greatest difference is
between users who are logged in versus users that clear their cookies.
Logged in users receive results that are reordered in two places (on average) as
compared to users with no cookies. Logged out users also receive reordered results compared
to the no cookie user, but the difference is smaller. The results in this figure \newtext{are consistent with the techniques that search engines are likely to use for personalization (i.e., per-user cookie tracking), and} give the first glimpse of how Google alters search results for different types of users.

The right column of Figure~\ref{fig:cookie} examines the impact of login cookies on Bing.
From the upper right figure (which plots the average Jaccard Index),
we see that, unlike Google Search, having Bing cookies does impact the results returned
from Bing. The lower right plot in Figure~\ref{fig:cookie} (which plots the average
Kendall's tau coefficient) demonstrates that cookies also influence the order of results from Bing.


We did not run our cookie-based experiments against DuckDuckGo because it does not place
cookies on users' browsers.


\begin{figure*}[t!]
        \includegraphics[width=\textwidth]{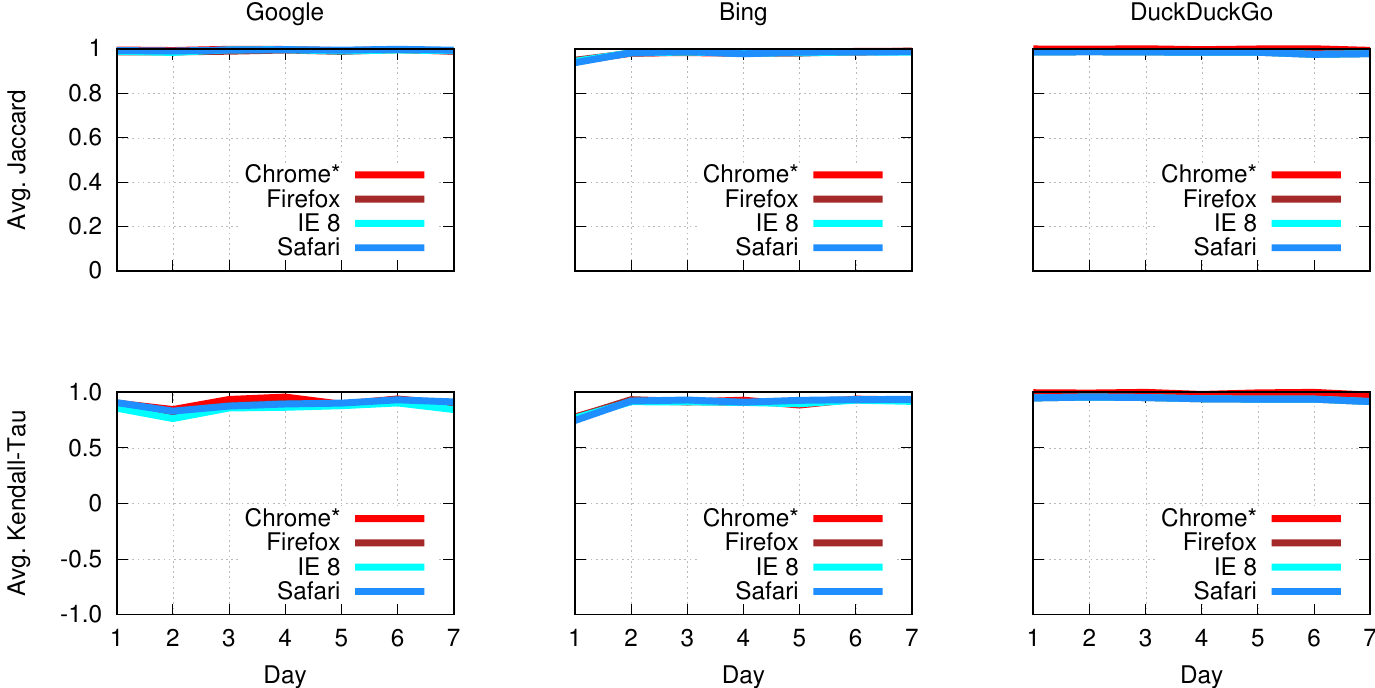}
        \caption{Results for the browser experiments on Google, Bing, and DuckDuckGo.}
        \label{fig:browser}
\end{figure*}

\paragraph{Browser User-Agent} Next, we examine whether the user's choice of browser
or Operating System (OS) can impact search results. To test this, we created 22
user accounts (11 for Google, 11 for Bing) and assigned
each one a different ``User-Agent'' string. As shown in Table~\ref{tab:features},
we encoded user-agents for 5 browsers and 4 OSs. Chrome 22 and
Windows 7 serve as the controls. For DuckDuckGo, we conduct the same experiment
sans user accounts, since DuckDuckGo does not have support for user accounts.

Figure~\ref{fig:browser} shows the results for our browser
experiments on Google, Bing, and DuckDuckGo. Unlike the cookie tracking experiment,
there is no clear differentiation between the different browsers
and the control experiment. The results for different OSs are similar, and we
omit them for brevity. Thus, we do not observe search personalization based on user-agent
strings for Google, Bing, or DuckDuckGo.


\paragraph{IP Address Geolocation} Next, we investigate whether the three
target search engines personalize results based on users' physical location.
To examine this, we create 22 user accounts (11 Google, 11 Bing) and run our
test suite while forwarding the traffic through
SSH tunnels to 10 geographically diverse PlanetLab machines. These PlanetLab
machines are located in the US states shown in Table~\ref{tab:features}.
Two accounts forward through the Massachusetts PlanetLab machine, since it is
the control. As before, we conduct this experiment against DuckDuckGo sans
user accounts.

\begin{figure*}[t!]
      \includegraphics[width=\textwidth]{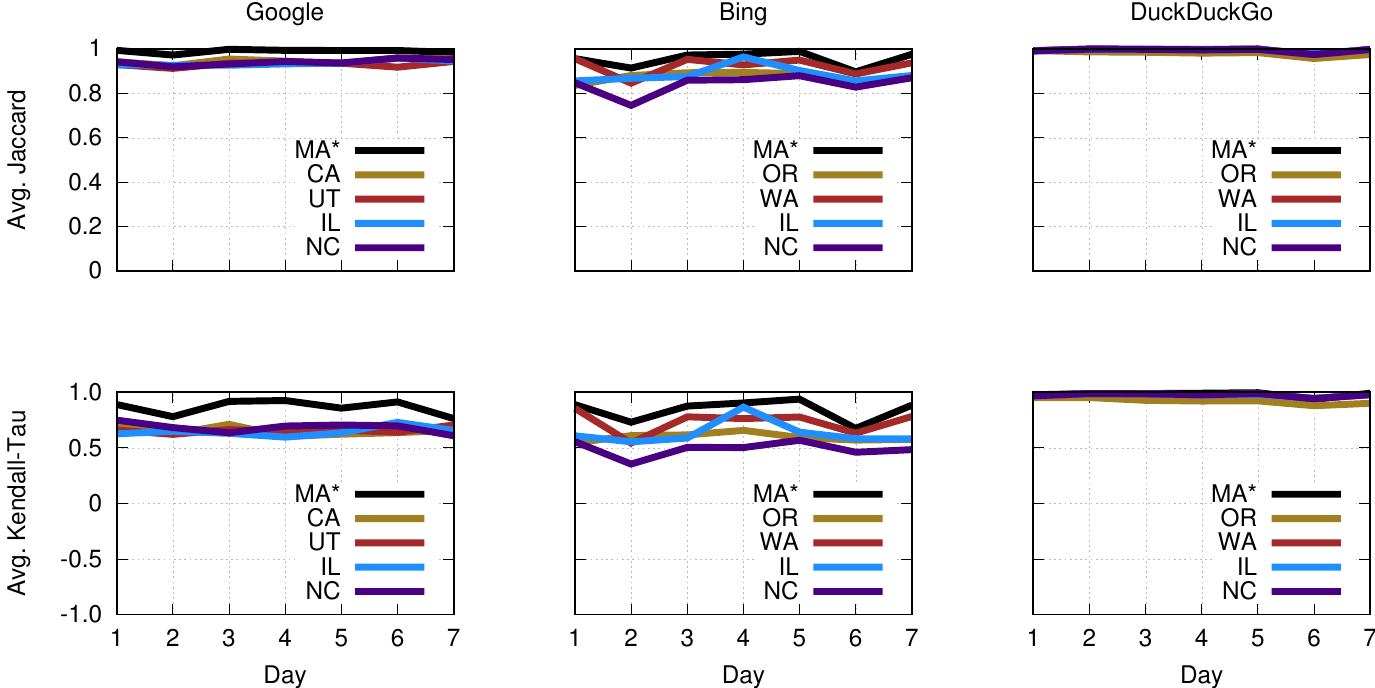}
  \caption{Results for the geolocation experiments on Google, Bing, and DuckDuckGo.}
  \label{fig:location}
\end{figure*}

Figure~\ref{fig:location} shows the results of our
location tests. There is a clear difference between the control and all the
other locations on both Google and Bing. On Google Search, the average
Jaccard Index for non-control tests is 0.91, meaning that queries from
different locations generally differ by one result. The same is true on
Bing, where the average Jaccard Index is 0.87. The difference
between locations is even more pronounced when we consider result order:
the average Kendall's tau coefficient for non-control accounts is 2.12 and 1.94 on
Google and Bing, respectively.

These results reveal that Google Search and Bing do personalize results based on
the user's geolocation. One example of this personalization can be seen by
comparing the MA and CA Google Search results for the query ``pier one'' (a home furnishing store).
The CA results include a link to a local news story covering a store grand
opening in the area. In contrast, the MA results include a Google Maps link
and a CitySearch link that highlight stores in the metropolitan area.

In contrast to Google and Bing, the search results from DuckDuckGo are essentially
identical regardless of the user's IP address. This result is not surprising, since it fits
with DuckDuckGo's stated policy of not personalizing search results for any reason.


\paragraph{Inferred Geolocation} During our experiments, we observed one set of anomalous results
from experiments that tunneled through Amazon EC2. In particular, 9 machines out
of 22 rented from Amazon's North Virginia datacenter were receiving heavily
personalized results, versus the other 13 machines, which showed no personalization.
Manual investigation revealed that Google Search was returning results with {\tt .co.uk}
links to the 9 machines, while the 13 other machines received zero {\tt .co.uk} links.
The 9 machines receiving UK results were all located in the same /16 subnet.

Although we could not determine
why Google Search believes the 9 machines are in the UK (we believe it is due to an incorrect
IP address geolocation database), we did confirm that this effect is independent of the Google account. 
As a result, we did not use EC2 machines as SSH tunnel endpoints for any of the results in this paper.
However, this anomoly does reveal that Google returns dramatically different search results
to users who are in different countries (or in this case, users Google believes are in different
countries).

\begin{figure*}[b!]
        \includegraphics[width=\textwidth]{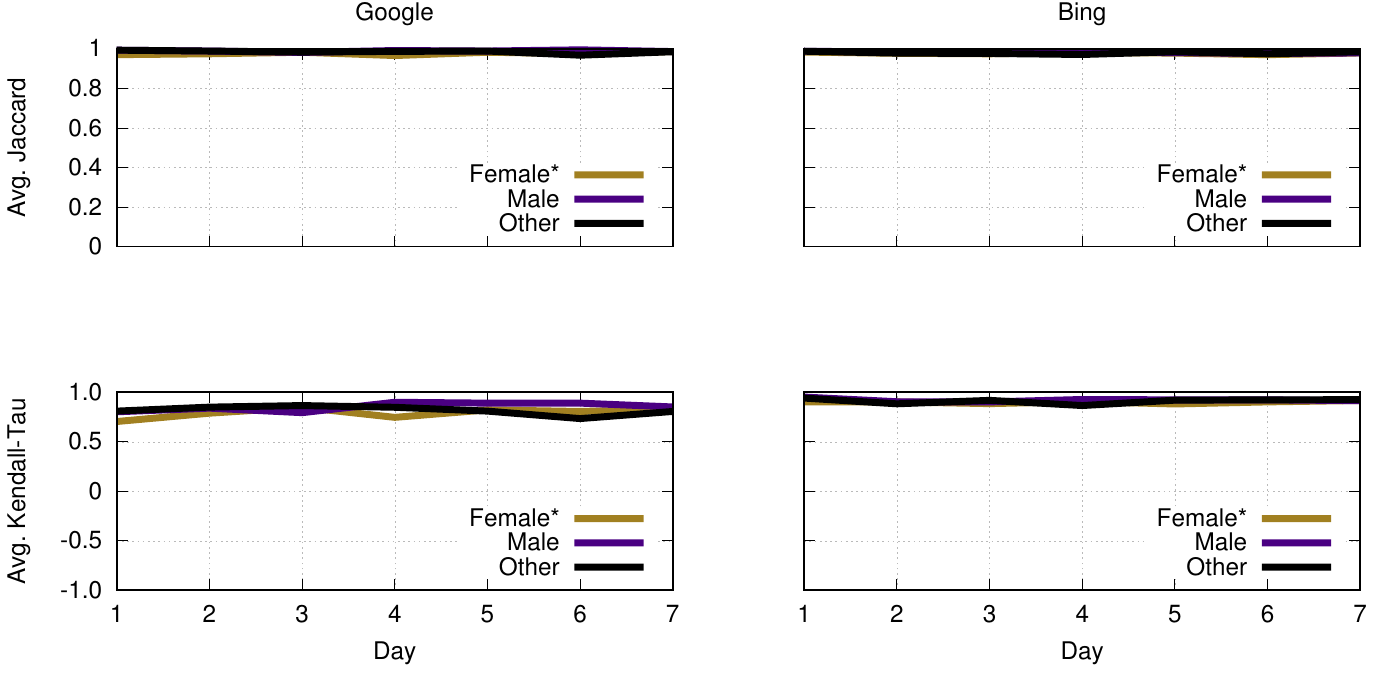}
        \caption{Results for the User Profile: Gender experiments on Google and Bing.}
        \label{fig:gender}
\end{figure*}


\paragraph{User Profile Attributes} In our next set of tests, we examine whether
Google Search and Bing uses demographic information from users' profiles to 
personalize results. Users must provide their gender and age when they sign up
for a Google account, which means that Google Search could leverage
this information to personalize results. Bing, on the other hand, collects
gender, age, and zip code.

To test this hypothesis, we created Google and Bing accounts with specific demographic
qualities. As shown in Table~\ref{tab:features}, we created ``female,'' ``male,''
and ``other'' accounts (these are the 3 choices Google and Bing give during account sign-up),
as well as accounts with ages 15 to 65, in increments of 10 years. On Bing, we
also create accounts from five different zip codes. The control
account in the gender tests is female, the control in the age tests is 15, and the
control in the zip code test is in Massachusetts.

The results for the gender test are presented in Figure~\ref{fig:gender}
We do not observe user profile gender-based personalization on Google or Bing.
Similarly, we do not observe personalization based on profile age
or zip code, and we omit the results for brevity. DuckDuckGo does not allow users
to create user accounts, so we do not run these tests on DuckDuckGo.

\subsection{Historical Features}

We now examine whether Google Search and Bing use an account's history of
activity to personalize results. We consider three types of historical
actions: prior searches, prior searches where the user clicks a result,
and Web browsing history.

To create a plausible series of actions for different accounts, we use data from Quantcast,
a Web analytics and advertising firm. Quantcast publishes a list of top websites (similar to
Alexa) that includes the {\em demographics} of visitors to sites~\cite{quantcast-2012},
broken down into the 20 categories shown in Table~\ref{tab:features}. 
Quantcast assigns each
website a score for each demographic, where scores $>$100 indicate that the given
demographic visits that website more frequently than average for the Web.
The larger the score, the more heavily weighted the site's visitors are towards a particular
demographic.

We use the Quantcast data to drive our historical experiments. In essence, our goal is to
have different accounts ``act'' like a member of each of Quantcast's demographic groups.
\newtext{The choice of our features was motivated by other web services and online advertisement services that 
use similar demographic categorizations to personalize content. 
Studies have shown that user demographics can be reliably
inferred from browsing histories, which can be useful for personalizing
content~\cite{hu-2007-predicting,goel-2012-icwsm}.}
Thus, for each of our 
experiments, we create 22 user accounts, two of which only run the 120 control queries, and 
20 of which perform actions (\ie searching, searching
and clicking, or Web browsing) based on their assigned demographic before running the 120
control queries. For example, one account builds Web browsing
history by visiting sites that are frequented by individuals earning $>$\$150k per year.
Each account is assigned a different Quantcast demographic, and chooses new action targets
each day using weighted random selection, where the weights are based on Quantcast
scores. For example, the $>$\$150k browsing history account chooses new sites to browse
each day from the corresponding list of URLs from Quantcast.

\begin{figure*}[t!]
        \includegraphics[width=\textwidth]{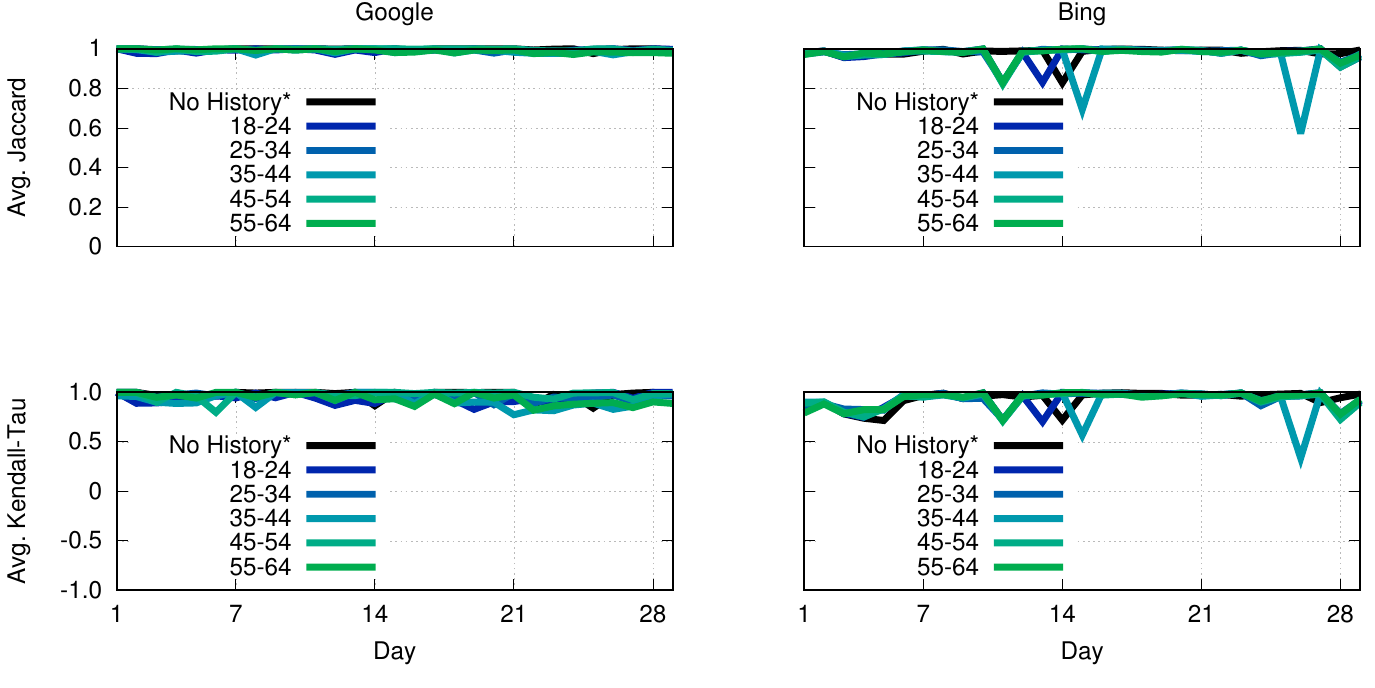}
        \caption{\newtext{Results for the Search History: Age Bracket experiments on Google and Bing.}}
        \label{fig:search_history}
\end{figure*}

We execute all three experimental treatments (searching, searching and clicking, and Web
browsing) on Google Search, but only execute two (searching, and searching and clicking) on
Bing. As previous studies have shown, Google is a ubiquitous presence across the
Web~\cite{roesner-2012-tracking}, which gives Google the ability to track user's as they
browse.
In contrast, Microsoft and Bing do not have a widespread presence: out of 1500 top sites
ranked by Quantcast, $<$1\% include cookies from Microsoft or its subsidiaries (\eg Live.com,
Outlook.com), versus 63\% for Google and its subsidiaries (\eg YouTube, Doubleclick).
Therefor, it is not feasible for Bing to track users' browsing behavior or personalize
search results based on browsing history.

DuckDuckGo does not use cookies, and thus has no way to track users or build up history.
Thus, we do not execute any of our historical experiments on DuckDuckGo.


\paragraph{Search History} First, we examine whether Google Search and/or Bing personalize
results based on search history. Each day, the 40 test accounts (20 for Google, 20 for Bing)
search for 100 demographic queries before executing the standard 120 queries.
The query strings are constructed by taking domains from the Quantcast top-2000
that have scores $>$100 for a particular demographic and removing subdomains and top level
domains (\eg {\tt www.amazon.com} becomes ``amazon'').

\newtext{Figure~\ref{fig:search_history} shows the results of the search history test for
five different age brackets. The ``No History'' account does not search for demographic
queries, and serves as the control. The vast majority of the time, all accounts receive
almost identical search results across both search engines (except for a few, random outliers
in the Bing results). If Google or Bing was personalizing search results based on search history,
we would expect the results for the age bracket accounts to diverge from the control results
over time. However, we do not observe this over the course of 30 days of experiments. This
observation holds for all of the demographic categories we tested, and we omit the results for
brevity. Thus, we do not observe personalization based on search history, although it is
possible that longer experiments could show larger differences.}

\paragraph{Search-Result-Click History} Next, we examine whether Google Search and/or Bing 
personalizes
results based on the search results that a user has clicked on. We use the same methodology
as for the search history experiment, with the addition that accounts click on the search
results that match their demographic queries. For example, an account that searches for
``amazon'' would click on a result linking to {\tt amazon.com}. Accounts will go through
multiple pages of search results to find the correct link for a given query.

The results of the click history experiments are the same as for the search history experiments.
There is little difference between the controls and the test accounts, regardless of demographic.
Thus, we do not observe personalization based on click history, and we omit the results for
brevity.

\paragraph{Browsing History} Next, we investigate whether Google Search personalizes results
based on Web browsing history (\ie by tracking users on third-party Web sites). 
In these experiments, each account logs into Google and then browses 5 random pages from 50
demographically skewed websites each day. We filter out websites that do not set Google 
cookies (or Google affiliates like DoubleClick), since Google cannot track visits
to these sites. Out of 1,587 unique domains in the Quantcast data that have scores $>$100, 700
include Google tracking cookies.

The results of the browsing history experiments are the same as for search history and
click history: regardless of demographic, we do not observe personalization. We omit these
results for brevity.

\begin{figure*}[t!]
        \includegraphics[width=\textwidth]{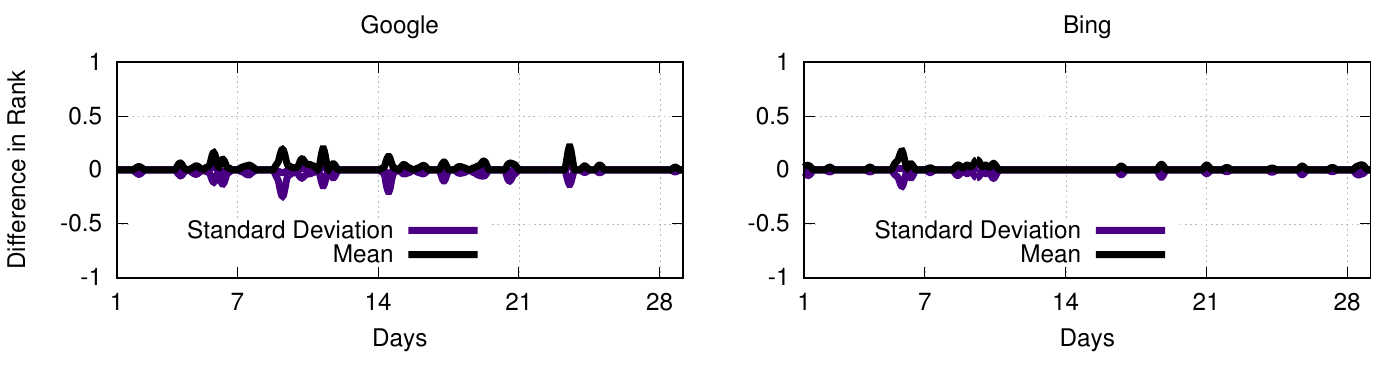}
        \caption{Results for the targeted domain clicking experiments on Google and Bing.}
        \label{fig:click_position}
\end{figure*}

\paragraph{Targeted Domain Clicking} Finally, we conduct a variant of our click history
experiment. In the previous search-result-click experiment, each account executed 100
``demographic'' searches and 120 standard test queries per day. However,
it is possible that this methodology is too complex to trigger search personalization,
\ie because each account creates such a diverse history of searches and clicks, the
search engines may have trouble isolating specific features to personalize on.

Thus, in this experiment, we simplify our methodology: we create 10 accounts, each of
which is assigned a specific, well-known news website. Each account executes 6 
news-related queries 4 times on each day (so, 24 searches each day, evenly spaced throughout
the day). After searching the account clicks on the link that is its assigned news website 
in the list of results. For example,
one account was assigned \url{www.foxnews.com}; 24 times per
day this account executed news-related queries, and always clicked
on results pointing to \url{www.foxnews.com} (if they appeared in the top 10 results).
In theory, this creates a very strong signal for personalization, \ie a search engine
could trivially observe that this user favors a specific website, and increase the
rank of results pointing to this website.

We conduct the targeted domain clicking test on both Google Search and Bing. We
created 10 experimental accounts on each search engine, each of which was assigned a
unique target domain, as well as 1 control account that searches but does not click
on any links.

Figure~\ref{fig:click_position} shows the results of our targeted domain clicking
experiments. To quantify our results, we plot the average difference in rank between
the targeted domains as seen by the experimental accounts and the control account. 
Difference of zero means that a particular domain appears at the same rank
for both the experimental account (which clicks on the domain) and the control
(which clicks on nothing). Positive difference in rank means the domain appears at
higher ranks for the experimental account, while a negative difference means that the
domain appears at higher ranks for the control.

As shown in Figure~\ref{fig:click_position}, on average, there is close to zero difference
between the ranks of domains, regardless of whether they have been clicked on. This
result holds true across Google Search and Bing. As shown by the standard deviation
lines, even when the rank of the domains differs, the variance is very low 
(\ie less than a single rank difference). Furthermore, although this test was run for
30 days, we do not observe divergence over time in the results; if the
search engines were personalizing results based on click history, we would expect
the difference in rank to increase over time as the experimental accounts build more
history. Thus, we conclude that clicking on results from particular domains does not
cause Google or Bing to elevate the rank of that domain.

\paragraph{Discussion} We were surprised that the history-driven tests did not reveal
personalization on Google Search or Bing. One explanation for this finding is that
account history may
only impact search results for a brief time window, \ie carry-over is the extent of
history-driven personalization on these search engines. 

\begin{figure*}[t!]
  \includegraphics[width=1.0\textwidth]{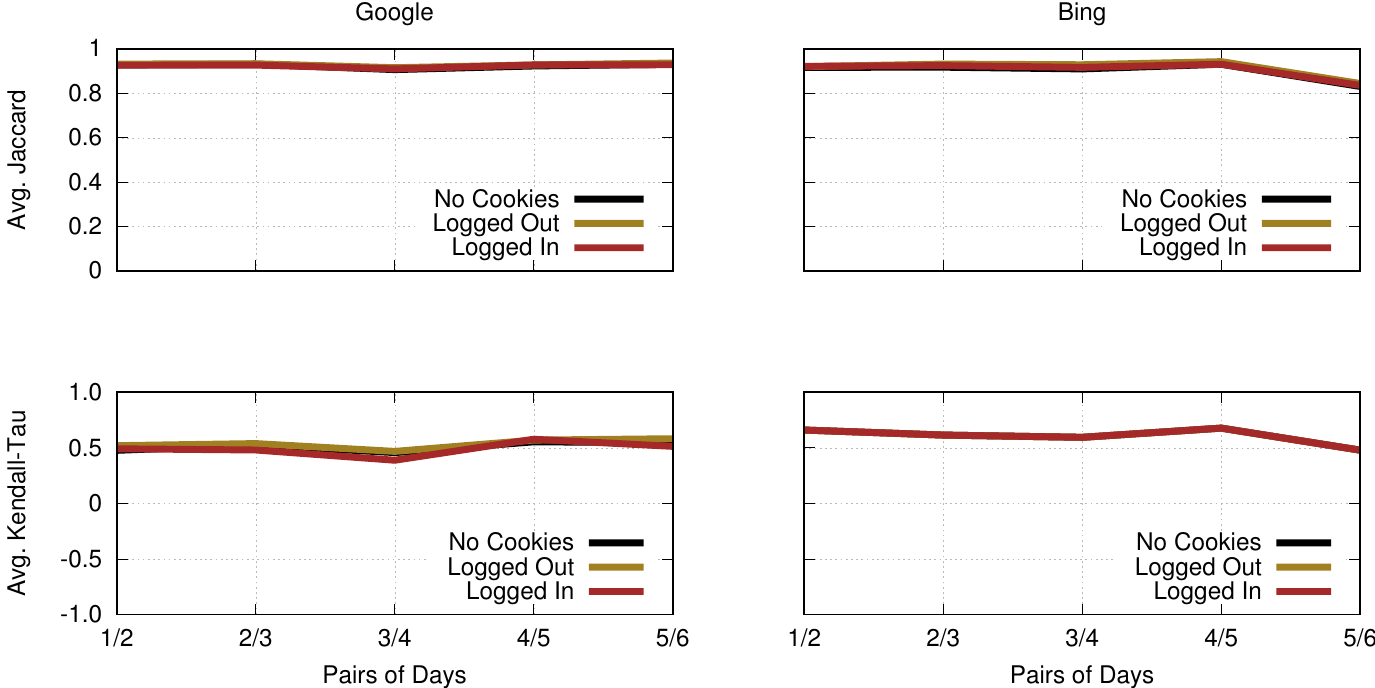}
  \caption{Day-to-day consistency of results for the cookie tracking experiments.}
  \label{fig:d2d}
\end{figure*}

\section{Quantifying Personalization}
\label{sec:practical}

In the previous section we demonstrate that Google Search personalization 
occurs based on 1) whether the user is logged in and 2) the location of the 
searching machine. In this section, we dive deeper into the data from our synthetic
experiments to better understand how personalization impacts search results. First,
we examine the temporal dynamics of search results. Next, we investigate the
amount of personalization in different categories of queries.
Finally, we examine the rank of personalized search results to understand whether
certain positions are more volatile than others.

\subsection{Temporal Dynamics}

In this section, we examine the temporal dynamics of results from Google Search and Bing
to understand how much search results change day-to-day, and whether
personalized results are more or less volatile than non-personalized search results.
To measure the dynamics of search engines over time, we compute the Jaccard
Index and Kendall Tau coefficient for search results from subsequent days. Figure~\ref{fig:d2d}
shows the day-to-day dynamics for our cookie tracking experiment (\ie the accounts
are logged in, logged out, and do not support cookies, respectively). The x-axis shows
which two days of search results are being compared, and each line corresponds to
a particular test account.

Figure~\ref{fig:d2d} reveals three facts about Google Search and Bing.
First, the lines in Figures~\ref{fig:d2d}
are roughly horizontal, indicating that the rate of change in the search indices
is roughly constant. Second, we observe that there is more reordering over time than new
results: average Jaccard Index on Google and Bing is $\approx$0.9, while average Kendall Tau coefficient
is 0.5 for Google and 0.7 for Bing.
Third, we observe that both of these trends are consistent across all
of our experiments, irrespective of whether the results are personalized.
This indicates that personalization does not increase the day-to-day volatility of
search results.

\begin{figure*}[t!]
  \includegraphics[width=1.0\textwidth]{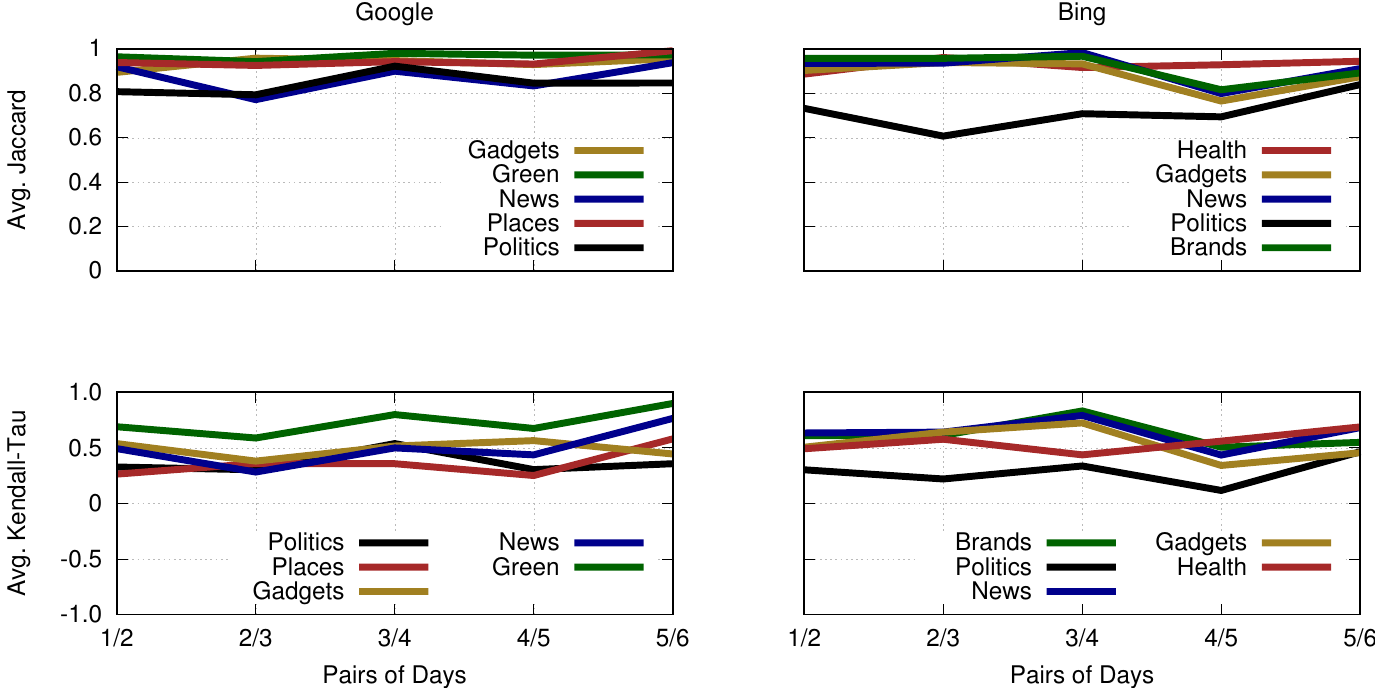}
  \caption{Day-to-day consistency within search query categories for the  cookie tracking experiments.}
  \label{fig:d2d_category}
\end{figure*}


\paragraph{Dynamics of Query Categories} We now examine the temporal dynamics of
results across different categories of queries. As shown in Table~\ref{tab:queries},
we use 12 categories of queries in our experiments. Our goal is to understand
whether each category is equally volatile over time, or whether certain categories
evolve more than others.

To understand the dynamics of query categories, we again calculate the
Jaccard Index and Kendall Tau coefficient between search results from subsequent days. However,
instead of grouping by experiment, we now group by query category.
Figure~\ref{fig:d2d_category} shows the day-to-day dynamics for query
categories during our cookie tracking experiments. Although we have 12 categories in total,
Figure~\ref{fig:d2d_category} only shows the 1 least volatile, and 4 most
volatile categories, for clarity. The results for all other experiments are
similar to the results for the cookie tracking test, and we omit them for brevity.

Figure~\ref{fig:d2d_category} reveals that the search results for different
query categories change at different rates day-to-day. For example, there are
more new results per day for ``politics'' related-queries on both Google Search and
Bing. Similarly, ``politics'' and ``gadgets'' related-queries both exhibit above
average reordering each day. This reflects how quickly information in these categories
changes on the Web.
In contrast, search queries in factual categories like ``what is'' and ``green''
(environmentally friendly topics) are less volatile over time (and are omitted
from Figure~\ref{fig:d2d_category} for clarity).

\subsection{Personalization of Query Categories}

We now examine the relationship between different categories of search queries
and personalization. In Section~\ref{sec:measure}, we demonstrate that Google Search
and Bing do personalize search results. However, it remains unclear whether all categories
of queries receive equal amounts of personalization.

\begin{figure*}[t]
  \centering \includegraphics[width=1.0\textwidth]{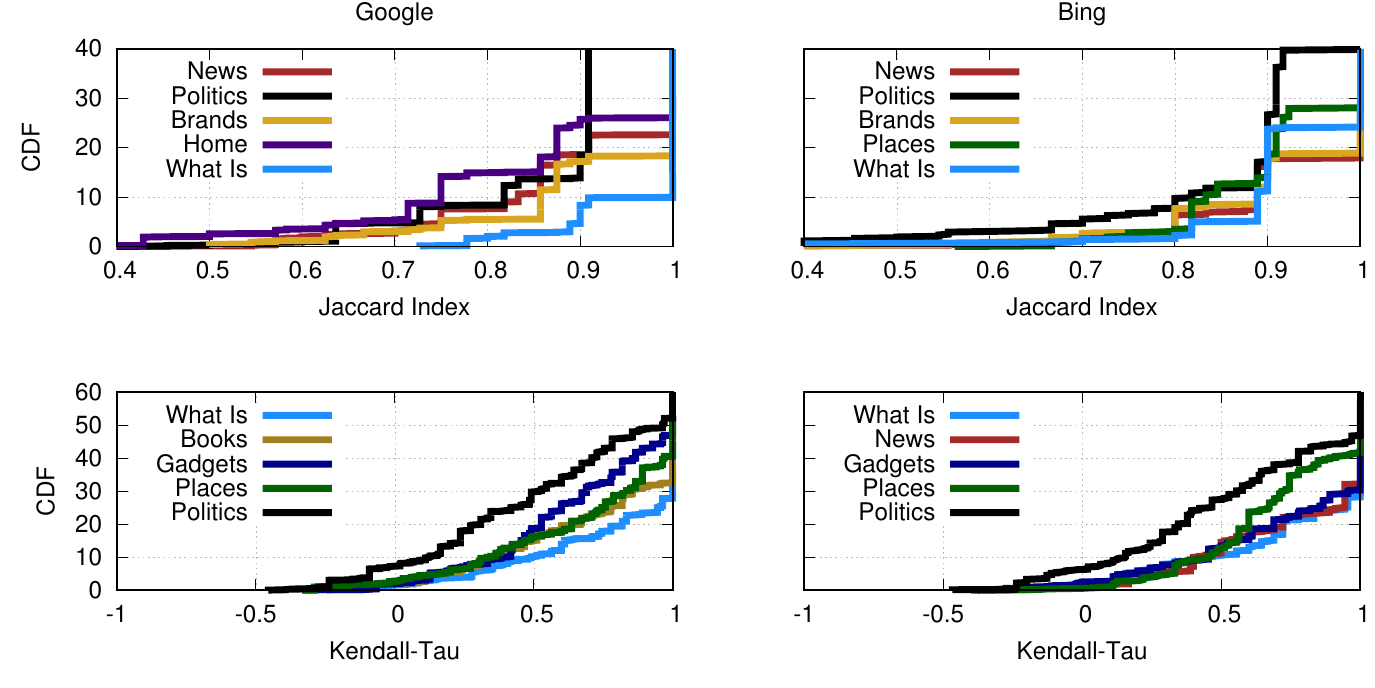}
  \caption{Differences in search results for five query categories on Google Search and Bing.}
  \label{fig:cdf_category}
\end{figure*}

To answer this question, we plot the cumulative distribute of Jaccard Index and Kendall Tau coefficient for
each category in Figure~\ref{fig:cdf_category}. These results are
calculated over all of our experiments (\ie User-Agent, Google Profile, geolocation,
{\em etc.}) for a single day of search results. For clarity, we only include lines
for the 1 most stable category (\ie Jaccard index and Kendall Tau are close to 1),
and the 4 least stable categories. 

Figure~\ref{fig:cdf_category} demonstrates that Google Search and Bing
personalize results for some query categories more than others. For example,
88\% of results for ``what is'' queries are identical on Google, while only 66\%
of results for ``gadgets'' are identical on Google. Overall, ``politics'' is the
most personalized query category on both search engines, followed by ``places''
and ``gadgets.'' CDFs calculated over other days of search results demonstrate
nearly identical results.

\subsection{Personalization and Result Ranking}

In this section, we focus on the volatility of results from Google Search and Bing at each
rank, with rank 1 being the first result on the page and rank 10 being the last result.
Understanding the impact of personalization on top ranked search results is critical,
since eye-tracking studies have demonstrated that users rarely scroll down
to results
``below the fold''~\cite{granka-2004-websearch,lorigo-2006-influence,guan-2007-websearch,cutrell-2007-websearch}.
Thus, we have two goals: 1) to understand whether certain ranks are more volatile in
general, and 2) to examine whether personalized search results are more volatile than
non-personalized results.

\begin{figure*}[t]
      \includegraphics[width=1.0\textwidth]{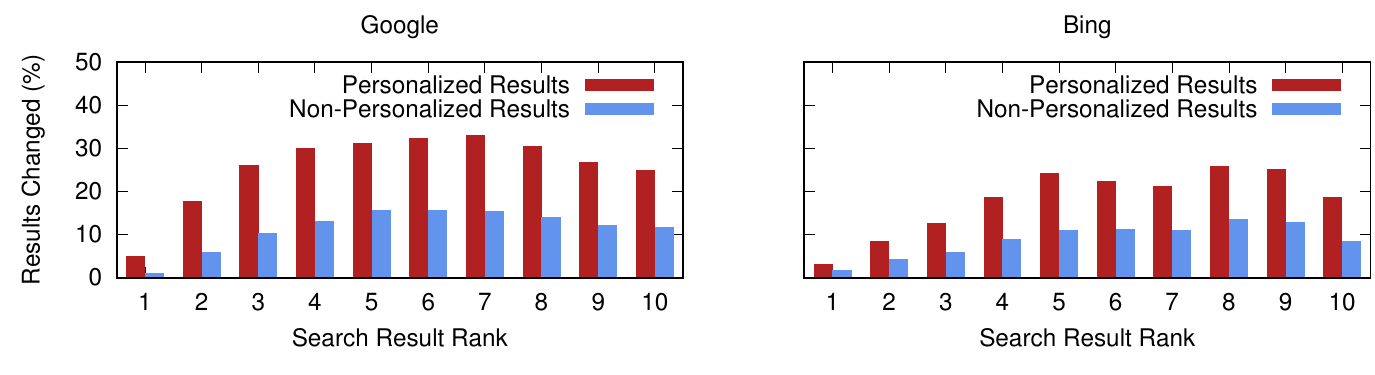}
  \caption{The percentage of results changed at each rank on Google Search and Bing.}
  \label{fig:rank}
\end{figure*}

To answer these questions, we plot Figure~\ref{fig:rank}, which shows the percentage
of results that change at each rank. To calculate these values, we perform a pairwise
comparison between the result at rank $r \in [1,10]$ received by a test account and the
corresponding control. We perform comparisons across all tests in
all experiments, across all seven days of measurement. This produces a total
number of results that are changed at each rank $r$, which we divide by the total
number of results at rank $r$ to produce a percentage. The personalized results
come from the cookie tracking and geolocation experiments;
all others experimental results are non-personalized.

Figure~\ref{fig:rank} reveals two interesting features. First, the results on
personalized pages are significantly more volatile than the results on non-personalized
pages. The result changes on non-personalized pages represent the noise floor of the
experiment; at nearly every rank, there are more than twice as many changes on personalized
pages. Second, Figure~\ref{fig:rank} shows that the volatility at each rank is not
uniform. Rank 1 exhibits the least volatility on Google Search and Bing. The volatility
increases until it peaks at 33\% in rank 7 on Google Search, and at 26\% in rank 8
on Bing. This indicates that both search engines are more conservative about
altering results at top ranks.

\begin{figure*}[b]
      \includegraphics[width=1.0\textwidth]{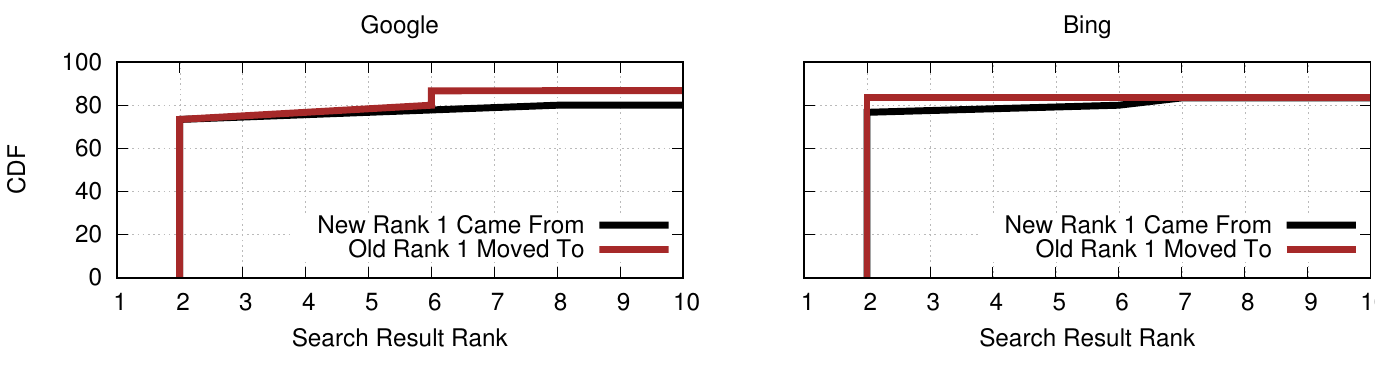}
  \caption{Movement of results to and from rank 1 for personalized searches.}
  \label{fig:resultone}
\end{figure*}

Given the extreme importance placed on rank 1 search results, we now delve deeper
into the rare cases where the result at rank 1 changes during personalized searches
(5\% of personalized rank 1 results change on Google, while 3\% change on Bing).
In each instance where the rank 1 result changes, we compare the results for the
test account and the control to determine 1) what was the {\em original rank} of the
result that moved to rank 1, and 2) what is the {\em new rank} of the result that used
to be at rank 1.

Figure~\ref{fig:resultone} plots the results of this test. In the
vast majority of cases, the rank 1 and 2 results switch places: on Google, 73\% of new rank
1 results originate from rank 2, and 58\% of old rank 1 results move to rank 2. 
On Bing, 77\% of new rank 1 results originate from rank 2, and 83\% of old rank 1 results
move to rank 2. Overall, on Google, 93\% of new rank 1 results come from the first
page of results, while 82\% of old rank 1 results remain somewhere on the first result page. 
On Bing, 83\% percent of rank 1 results come from or move to somewhere on the first page of
results. However, none of the CDFs sum to 100\%, \ie there are cases where the new rank 1
result does not appear in the control results and/or the old rank 1 result disappears
completely from the test results. The latter case is more common on Google, with 18\% of rank 1
results getting evicted completely from the first page of results. Both cases, are equally
likely on Bing.

Figure~\ref{fig:resultone} reveals similarities and differences between Google
Search and Bing. On one hand, both search engines are clearly conservative
about changing rank 1 search results, \ie the vast majority of changes are simply swaps between
rank 1 and 2. On the other hand, when the rank 1 result does change, Google and Bing leverage
different strategies: Google Search prefers to elevate results already on the first page,
while Bing prefers to insert completely new links. We manually examined instances where
Bing inserted new results at rank 1, and found that in most cases these new links were to
Bing services, \eg a box of links Bing News results.

\begin{figure*}[t]
      \includegraphics[width=1.0\textwidth]{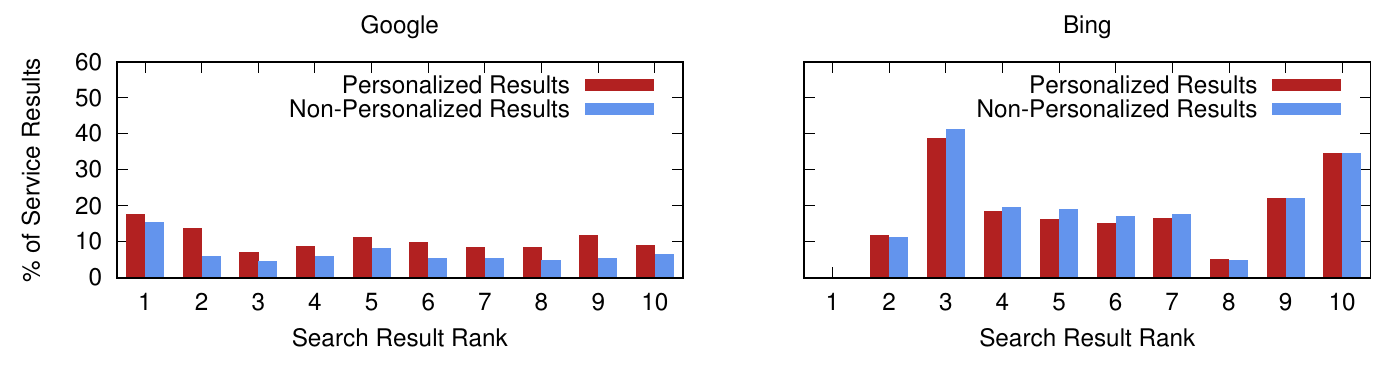}
  \caption{Rank of embedded services in search results from Google and Bing.}
  \label{fig:service_rank}
\end{figure*}

\newtext{
\subsection{Personalization and Aggregated Search}
}
In Section~\ref{sec:terminology}, we noted that some of the results from search engines
do not point to third-party websites. Instead, some results embed links and functionality
from {\em services} maintained by the search engine provider. \newtext{The inclusion of links to 
other services in search results is sometimes referred to as ``aggregated search.''}
For example, Google often embeds results from Google News,
Google Maps, YouTube, and Google+ into pages of search results. Figure~\ref{fig:google-example}
shows an example of an embedded Google service: a box of queries that are ``Related'' to
the given search query. Bing offers an array of similar services and embeds to them in
search results.

In this section, we examine links to provider services in search results. Specifically,
we are interested in whether personalization impacts the placement and amount of links
to provider services. These are important questions, given that regulators have questioned
the placement of provider services in search results within the context of antitrust
regulation~\cite{robinson-2013-googvseu}, \ie do providers promote their own
services at the expense of third-party websites?

First, we examine the percentage of results at each rank that embed provider services.
Figure~\ref{fig:service_rank} shows the percentage of results at each rank that embed
provider services on Google and Bing. We split our data into personalized and non-personalized
pages of results, where results from the logged-in/out and location experiments are considered
to be personalized. We aggregate across all 120 experimental queries and all 30 days of experiments.

Figure~\ref{fig:service_rank} demonstrates that Google and Bing exhibit different behavior
when it comes to embedding provider services. Overall, Bing embeds its
own services 19\% of the time, whereas Google embeds its own services 9\% of the time.
However, Google embeds services at rank 1 on $\approx$15\% of result pages, whereas Bing
{\em never} embeds services at rank 1. Instead, Google tends to embed services
uniformly across ranks 2-10, whereas Bing favors embedding services at ranks 3 and 10.
On Bing, the rank 3 results point to a variety of different services (\eg Bing Images,
Bing Video, Bing News, {\em etc.}), while the service at rank 10 is almost always
Related Searches.

Figure~\ref{fig:service_rank} also shows that personalization only appears to influence
service embedding on Google. On Bing, the amount and placement of embedded services does
not change between personalized and non-personalized search results. However, on Google,
12\% of links on personalized pages point to services, versus 8\% on
non-personalized pages. This trend is relatively uniform across all ranks on Google.
This demonstrates that personalization does increase the number of embedded
services seen by Google Search users.

\begin{figure*}[b]
      \includegraphics[width=1.0\textwidth]{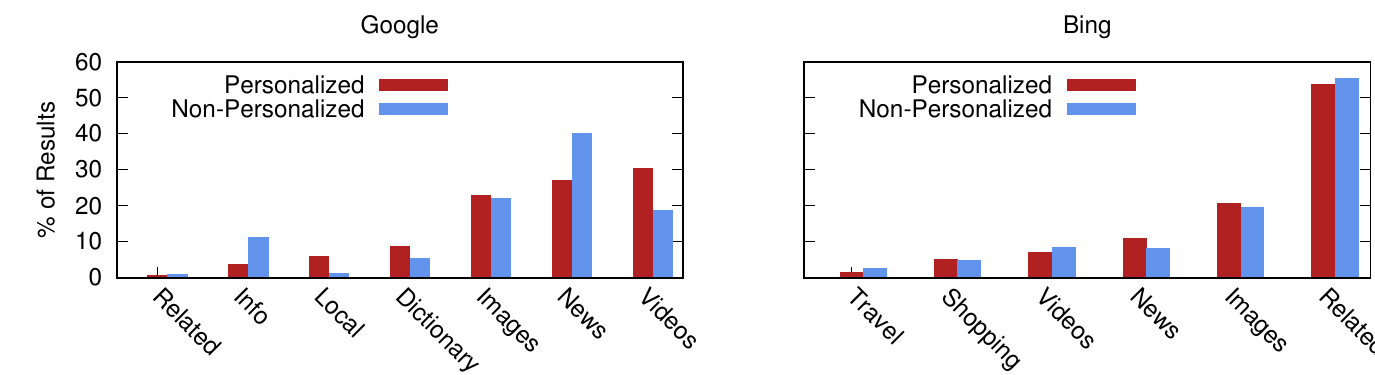}
  \caption{Percentage of embeddings of different services on Google and Bing.}
  \label{fig:service_type}
\end{figure*}

Next, we seek to understand whether personalization impacts which services are embedded
by search engines. To answer this question, we plot Figure~\ref{fig:service_type},
which shows the percentage of results that can be attributed to different services
(we only consider links to services, so the bars sum to 100\%). As before, we examine
results on personalized and non-personalized pages separately, and aggregate across
all 120 search queries and all 30 days of experiments.

On Google, the top three most embedded services are Google Images, Google News, and Google
Video (which also includes links to YouTube). Bing also generates many links to its
equivalent image, news, and video services, however the most embedded service by a
large margin is Related Searches (which is almost always placed at rank 10).
In contrast, Google only embeds Related Searches into 1\% of results pages.

Figure~\ref{fig:service_type} reveals that Google does personalize the types of
services it embeds, whereas Bing does not. On Google, ``Info'' and Google News results
tend to be served on more non-personalized pages. Info results present information from
Google's Knowledge Graph, which are usually answers to questions, \eg "Madrid" if the
user searches for "Capital of Spain." Conversely, ``Local,'' Dictionary, and Google Video
results are served more frequently on personalized pages. Local results present lists
of geographically close stores and restaurants that are related to the user's query,
\eg a list of pizza parlors if the user searches for ``pizza.'' In contrast to Google,
Bing embeds different services at roughly static rates, regardless of personalization.

\paragraph{Google and Bing News} As shown in Figure~\ref{fig:service_type}, pages of search
results from Google and
Bing often include embedded boxes of news from their respective news portals (Google News
and Bing News). Each box is characterized by 1-4 links to news stories on third-party websites
(\eg CNN, New York Times, Fox News, {\em etc.}) that are relevant to the given query.
Typically, one or more of the links will be enhanced with a thumbnail image also taken
from a third-party website.

It is known that services like Google News use personalization to recommend
news stories that the service believes are relevant to each user~\cite{das-2007-news}.
This raises the question: are the news links embedded in pages of search results also personalized?
To answer this question, we went through our dataset and located all instances where an
experimental account and the corresponding control account were both served embedded news
boxes. In these cases, we compared the links in each embedded news box to examine
whether news results are being personalized.

The results of this experiment show that search results from Google News and Bing News
are not personalized, even if other results on the page are personalized. On both Google
and Bing, the Jaccard Index when comparing news links from experimental and control accounts
is consistently $\approx$1, and the Kendall Tau coefficient is consistently $\approx$1. These results
are the same regardless of the characteristics of the experimental account (\ie location,
logged in/out, search history, {\em etc.}). Thus, it appears that Google and Bing only
personalize news directly on their respective news sites, but not in their search results.

\section{Concluding Discussion}
\label{sec:conclusion}

Over the past few years, we have witnessed a trend of personalization in
numerous Internet-based services, including Web search. While
personalization provides obvious benefits for users, it also opens up the
possibility that certain information may be unintentionally hidden from users.
Despite the variety of speculation on this topic, to date, there has been little
quantification of the basis and extent of personalization in Web search services today.

In this paper, we introduce a robust methodology for measuring personalization on Web
search engines. Our
methodology controls for numerous sources of noise, allowing us to accurately measure
the extent of personalization. We applied our methodology to real Google and Bing
accounts recruited from AMT and observe that 11.7\% of search results on Google and 
15.8\% on Bing show
differences due to  personalization. Using artificially created accounts, we observe
that measurable personalization on Google Search and Bing \newtext{is triggered by} 1) being
logged in to a user account and 2) making requests from different geographic areas.

However, much work remains to be done: we view our results
as a first step in providing transparency for users of Web search and other Web-based
services. In the paragraphs below, we discuss a few of the issues brought up by our work,
as well as promising directions for future research.

\paragraph{Scope} In this paper, we focus on queries to US versions of the Google Web Search,
Bing, and DuckDuckGo.
All queries are in English, and are drawn from topics that are primarily of interest to
US residents. We leave the examination of search engines in other countries and other
languages to future work.

\paragraph{Incompleteness} \newtext{As a result of our methodology, we are only able to identify positive
instances of personalization; we cannot claim the absence of personalization, as we
may not have considered other dimensions along which personalization could occur and we can only 
test a finite set of search terms. However, the
dimensions that we chose to examine in this paper are the most obvious ones for personalization
(considering how much prior work has looked at demographic, location-based, and history-based
personalization). An interesting extension of our experiments would be to determine how the 
popularity of the query terms effect personalization.}

Given that any form of personalization is a moving target, we aim to
continue this work by looking at additional categories of Web searches,
examining searches from mobile devices, and looking at other user behaviors (\eg using
services like Gmail, Google+, and Google Maps). We also plan on examining the impact
of mechanisms that may disable personalization (\eg enabling Do-Not-Track headers).

\paragraph{Generality} The methodology that we develop is not specific to Google Web Search,
Bing, or DuckDuckGo.
The sources of noise that we control for are present in other search engines (\eg
Google News Search) as well as other Web-based services (\eg Twitter search, Yelp
recommendations, {\em etc.}). We plan on applying our methodology to these and other
search services to quantify personalization of different types.

\newtext{\paragraph{Recommendations} We have focused on measuring the personalization that exists on 
popular Web search engines today, and have avoided trying to determine whether the personalization we 
observed resulted in better or worse search experiences for the users.  We leave such an exploration 
to future work.  However, we do believe that services today should make personalization transparent 
to the users (\ie label results that are personalized as such) and allow users to disable personalization 
if they wish.}

\paragraph{Impact} In this paper, we focused on quantifying literal differences in
search results, \eg {\tt a.com} is different from {\tt b.com}. However, we do not
address the issue of semantic differences, \ie
do {\tt a.com} and {\tt b.com} contain different information content? If so, what
is the impact of these differences?  While semantic differences and impact are challenging to
quantify, we plan to explore natural language processing and user studies as a first step.

\paragraph{Open Source} We make all of the crawling and parsing code, as well as the Google Web
Search, Bing, and DuckDuckGo data from Section~\ref{sec:measure}, available to the research
community at
\begin{center}
\vspace{-0.5mm}
{\tt http://personalization.ccs.neu.edu/}
\vspace{-0.5mm}
\end{center}

\bibliographystyle{ACM-Reference-Format}
\bibliography{all}


\begin{thebibliography}{00}


\ifx \showCODEN    \undefined \def \showCODEN     #1{\unskip}     \fi
\ifx \showDOI      \undefined \def \showDOI       #1{{\tt DOI:}\penalty0{#1}\ }
  \fi
\ifx \showISBNx    \undefined \def \showISBNx     #1{\unskip}     \fi
\ifx \showISBNxiii \undefined \def \showISBNxiii  #1{\unskip}     \fi
\ifx \showISSN     \undefined \def \showISSN      #1{\unskip}     \fi
\ifx \showLCCN     \undefined \def \showLCCN      #1{\unskip}     \fi
\ifx \shownote     \undefined \def \shownote      #1{#1}          \fi
\ifx \showarticletitle \undefined \def \showarticletitle #1{#1}   \fi
\ifx \showURL      \undefined \def \showURL       #1{#1}          \fi
\providecommand\bibfield[2]{#2}
\providecommand\bibinfo[2]{#2}
\providecommand\natexlab[1]{#1}
\providecommand\showeprint[2][]{arXiv:#2}

\bibitem[\protect\citeauthoryear{Andersen, Giversen, Jensen, Larsen, Pedersen,
  and Skyt}{Andersen et~al\mbox{.}}{2000}]%
        {andersen-2000-clickstreams}
\bibfield{author}{\bibinfo{person}{Jesper Andersen}, \bibinfo{person}{Anders
  Giversen}, \bibinfo{person}{Allan~H. Jensen}, \bibinfo{person}{Rune~S.
  Larsen}, \bibinfo{person}{Torben~Bach Pedersen}, {and} \bibinfo{person}{Janne
  Skyt}.} \bibinfo{year}{2000}\natexlab{}.
\newblock \showarticletitle{Analyzing Clickstreams Using Subsessions}, In
  \bibinfo{booktitle}{ACM International Workshop On Data Warehousing and OLAP}.
  \bibinfo{journal}{{\em ACM International Workshop On Data Warehousing and
  OLAP\/}} (\bibinfo{date}{November} \bibinfo{year}{2000}),
  \bibinfo{pages}{25--32}.
\newblock


\bibitem[\protect\citeauthoryear{Andrade and Silva}{Andrade and Silva}{2006}]%
        {andrade-2006-relevance}
\bibfield{author}{\bibinfo{person}{Leonardo Andrade} {and}
  \bibinfo{person}{M\'{a}rio~J. Silva}.} \bibinfo{year}{2006}\natexlab{}.
\newblock \showarticletitle{Relevance Ranking for Geographic IR}. In
  \bibinfo{booktitle}{{\em ACM Workshop On Geographic Information Retrieval}}.
  \bibinfo{address}{Seattle, Washington}.
\newblock


\bibitem[\protect\citeauthoryear{Bar-Ilan, Keenoy, Yaari, and Levene}{Bar-Ilan
  et~al\mbox{.}}{2007}]%
        {barilan-2007-rankings}
\bibfield{author}{\bibinfo{person}{Judit Bar-Ilan}, \bibinfo{person}{Kevin
  Keenoy}, \bibinfo{person}{Eti Yaari}, {and} \bibinfo{person}{Mark Levene}.}
  \bibinfo{year}{2007}\natexlab{}.
\newblock \showarticletitle{User rankings of search engine results}.
\newblock \bibinfo{journal}{{\em Journal of the American Society for
  Information Science and Technology\/}} \bibinfo{volume}{58},
  \bibinfo{number}{9} (\bibinfo{date}{July} \bibinfo{year}{2007}),
  \bibinfo{pages}{1254--1266}.
\newblock


\bibitem[\protect\citeauthoryear{Bing It On}{Bing It On}{}]%
        {BingItOn}
Bing It On.
\newblock \bibinfo{title}{Bing It on}.
\newblock   (\bibinfo{year}{????}).
\newblock
\newblock
\shownote{\url{http://www.bingiton.com}.}


\bibitem[\protect\citeauthoryear{Broder}{Broder}{2002}]%
        {broder-2002-taxonomy}
\bibfield{author}{\bibinfo{person}{Andrei Broder}.}
  \bibinfo{year}{2002}\natexlab{}.
\newblock \showarticletitle{A Taxonomy of Web Search}.
\newblock \bibinfo{journal}{{\em ACM SIGIR Forum\/}} \bibinfo{volume}{36},
  \bibinfo{number}{2} (\bibinfo{year}{2002}), \bibinfo{pages}{3--10}.
\newblock


\bibitem[\protect\citeauthoryear{Chan, Kumar, Ma, and Koehler}{Chan
  et~al\mbox{.}}{2012}]%
        {chan-2012-organic}
\bibfield{author}{\bibinfo{person}{David Chan}, \bibinfo{person}{Deepak Kumar},
  \bibinfo{person}{Sheng Ma}, {and} \bibinfo{person}{Jim Koehler}.}
  \bibinfo{year}{2012}\natexlab{}.
\newblock \bibinfo{booktitle}{{\em Impact of Ranking of Organic Search Results
  on the Incrementality of Search Ads}}.
\newblock \bibinfo{type}{{T}echnical {R}eport}. \bibinfo{institution}{Google}.
\newblock


\bibitem[\protect\citeauthoryear{Cheng, Gao, and Liu}{Cheng
  et~al\mbox{.}}{2010}]%
        {cheng-2010-intent}
\bibfield{author}{\bibinfo{person}{Zhicong Cheng}, \bibinfo{person}{Bin Gao},
  {and} \bibinfo{person}{Tie-Yan Liu}.} \bibinfo{year}{2010}\natexlab{}.
\newblock \showarticletitle{Actively Predicting Diverse Search Intent from User
  Browsing Behaviors}. In \bibinfo{booktitle}{{\em International World Wide Web
  Conference}}. \bibinfo{address}{Raleigh, North Carolina}.
\newblock


\bibitem[\protect\citeauthoryear{comScore August 2012 U.S. Search
  Results}{comScore August 2012 U.S. Search Results}{2012}]%
        {GoogleSearchStatistics}
comScore August 2012 U.S. Search Results \bibinfo{year}{2012}\natexlab{}.
\newblock \bibinfo{title}{comScore August 2012 U.S. Search Results}.
\newblock   (\bibinfo{year}{2012}).
\newblock
\newblock
\shownote{\url{http://bit.ly/ThGnOc}.}


\bibitem[\protect\citeauthoryear{Craver}{Craver}{2013}]%
        {craver-2013-bingsocial}
\bibfield{author}{\bibinfo{person}{Thom Craver}.}
  \bibinfo{year}{2013}\natexlab{}.
\newblock \bibinfo{title}{Bing Further Bolsters Social Results with 5x More
  Facebook Content}.
\newblock   (\bibinfo{date}{January} \bibinfo{year}{2013}).
\newblock
\newblock
\shownote{\url{http://bit.ly/Wnm97D}.}


\bibitem[\protect\citeauthoryear{Crook}{Crook}{2011}]%
        {crook-2011-bingadaptive}
\bibfield{author}{\bibinfo{person}{Aidan Crook}.}
  \bibinfo{year}{2011}\natexlab{}.
\newblock \bibinfo{title}{Adapting Search to You}.
\newblock   (\bibinfo{date}{September} \bibinfo{year}{2011}).
\newblock
\newblock
\shownote{\url{http://binged.it/10MWc6j}.}


\bibitem[\protect\citeauthoryear{Crown and Nelson}{Crown and Nelson}{2007}]%
        {crown-2007-disagree}
\bibfield{author}{\bibinfo{person}{Frank Crown} {and}
  \bibinfo{person}{Michael~L. Nelson}.} \bibinfo{year}{2007}\natexlab{}.
\newblock \showarticletitle{Agreeing to Disagree: Search Engines and Their
  Public Interfaces}. In \bibinfo{booktitle}{{\em ACM/IEEE-CS Joint Conference
  on Digital Libraries}}. \bibinfo{address}{Vancouver, Canada}.
\newblock


\bibitem[\protect\citeauthoryear{Cutrell and Guan}{Cutrell and Guan}{2007}]%
        {cutrell-2007-websearch}
\bibfield{author}{\bibinfo{person}{Edward Cutrell} {and}
  \bibinfo{person}{Zhiwei Guan}.} \bibinfo{year}{2007}\natexlab{}.
\newblock \showarticletitle{What are you looking for?: an eye-tracking study of
  information usage in web search}. In \bibinfo{booktitle}{{\em Annual
  Conference of the ACM Special Interest Group on Computer Human Interaction}}.
  \bibinfo{address}{San Jose, California}.
\newblock


\bibitem[\protect\citeauthoryear{Das, Datar, Garg, and Rajaram}{Das
  et~al\mbox{.}}{2007}]%
        {das-2007-news}
\bibfield{author}{\bibinfo{person}{Abhinandan Das}, \bibinfo{person}{Mayur
  Datar}, \bibinfo{person}{Ashutosh Garg}, {and} \bibinfo{person}{Shyam
  Rajaram}.} \bibinfo{year}{2007}\natexlab{}.
\newblock \showarticletitle{Google News Personalization: Scalable Online
  Collaborative Filtering}. In \bibinfo{booktitle}{{\em International World
  Wide Web Conference}}. \bibinfo{address}{Banff, Canada}.
\newblock


\bibitem[\protect\citeauthoryear{Dou, Song, and Wen}{Dou et~al\mbox{.}}{2007}]%
        {dou-2007-personalized}
\bibfield{author}{\bibinfo{person}{Zhicheng Dou}, \bibinfo{person}{Ruihua
  Song}, {and} \bibinfo{person}{Ji-Rong Wen}.} \bibinfo{year}{2007}\natexlab{}.
\newblock \showarticletitle{A Large-scale Evaluation and Analysis of
  Personalized Search Strategies}. In \bibinfo{booktitle}{{\em International
  World Wide Web Conference}}. \bibinfo{address}{Banff, Canada}.
\newblock


\bibitem[\protect\citeauthoryear{Fan, Gordon, and Pathak}{Fan
  et~al\mbox{.}}{2000}]%
        {fan-2000-personalization}
\bibfield{author}{\bibinfo{person}{Weiguo Fan}, \bibinfo{person}{Michael~D.
  Gordon}, {and} \bibinfo{person}{Praveen Pathak}.}
  \bibinfo{year}{2000}\natexlab{}.
\newblock \showarticletitle{Personalization of Search Engine Services for
  Effective Retrieval and Knowledge Management}. In \bibinfo{booktitle}{{\em
  Annual Conference on Information Systems}}. \bibinfo{address}{Atlanta,
  Georgia}.
\newblock


\bibitem[\protect\citeauthoryear{Farber}{Farber}{2013}]%
        {google-brain-2013}
\bibfield{author}{\bibinfo{person}{Dan Farber}.}
  \bibinfo{year}{2013}\natexlab{}.
\newblock \bibinfo{title}{Google Search scratches its brain 500 million times a
  day}.
\newblock   (\bibinfo{date}{May} \bibinfo{year}{2013}).
\newblock
\newblock
\shownote{\url{http://cnet.co/127g56f}.}


\bibitem[\protect\citeauthoryear{Gauch, Chaffee, and Pretschner}{Gauch
  et~al\mbox{.}}{2003}]%
        {gauch-2003-ontologybased}
\bibfield{author}{\bibinfo{person}{Susan Gauch}, \bibinfo{person}{Jason
  Chaffee}, {and} \bibinfo{person}{Alexander Pretschner}.}
  \bibinfo{year}{2003}\natexlab{}.
\newblock \showarticletitle{Ontology-based personalized search and browsing}.
\newblock \bibinfo{journal}{{\em Web Intelligence and Agent Systems\/}}
  \bibinfo{volume}{1} (\bibinfo{year}{2003}), \bibinfo{pages}{1--3}.
\newblock


\bibitem[\protect\citeauthoryear{Goel, Hofman, and Sirer}{Goel
  et~al\mbox{.}}{2012}]%
        {goel-2012-icwsm}
\bibfield{author}{\bibinfo{person}{Sharad Goel}, \bibinfo{person}{Jake~M.
  Hofman}, {and} \bibinfo{person}{M~Irmak Sirer}.}
  \bibinfo{year}{2012}\natexlab{}.
\newblock \showarticletitle{Who Does What on the Web: A Large-scale Study of
  Browsing Behavior}. In \bibinfo{booktitle}{{\em International AAAI Conference
  on Weblogs and Social Media}}. \bibinfo{address}{Dublin, Ireland}.
\newblock


\bibitem[\protect\citeauthoryear{Google}{Google}{2005}]%
        {psearch-2005}
\bibfield{author}{\bibinfo{person}{Google}.} \bibinfo{year}{2005}\natexlab{}.
\newblock \bibinfo{title}{Personalized Search Graduates from Google Labs}.
\newblock   (\bibinfo{date}{November} \bibinfo{year}{2005}).
\newblock
\newblock
\shownote{\url{http://googlepress.blogspot.com/2005/11/personalized-search-graduates-from_10.html}.}


\bibitem[\protect\citeauthoryear{Google Zeitgeist}{Google Zeitgeist}{2012}]%
        {GoogleZeitgeist}
Google Zeitgeist \bibinfo{year}{2012}\natexlab{}.
\newblock \bibinfo{title}{Google Zeitgeist}.
\newblock   (\bibinfo{year}{2012}).
\newblock
\newblock
\shownote{\url{http://www.googlezeitgeist.com}.}


\bibitem[\protect\citeauthoryear{Granka, Joachims, and Gay}{Granka
  et~al\mbox{.}}{2004}]%
        {granka-2004-websearch}
\bibfield{author}{\bibinfo{person}{Laura~A. Granka}, \bibinfo{person}{Thorsten
  Joachims}, {and} \bibinfo{person}{Geri Gay}.}
  \bibinfo{year}{2004}\natexlab{}.
\newblock \showarticletitle{Eye-tracking analysis of user behavior in WWW
  search}. In \bibinfo{booktitle}{{\em Conference of the ACM Special Interest
  Group on Information Retrieval}}. \bibinfo{address}{Sheffield, United
  Kingdom}.
\newblock


\bibitem[\protect\citeauthoryear{Green}{Green}{2011}]%
        {green-2011-filterbubble}
\bibfield{author}{\bibinfo{person}{Holly Green}.}
  \bibinfo{year}{2011}\natexlab{}.
\newblock \bibinfo{title}{Breaking Out of Your Internet Filter Bubble}.
\newblock   (\bibinfo{date}{August} \bibinfo{year}{2011}).
\newblock
\newblock
\shownote{\url{http://onforb.es/oYWbDf}.}


\bibitem[\protect\citeauthoryear{Guan and Cutrell}{Guan and Cutrell}{2007}]%
        {guan-2007-websearch}
\bibfield{author}{\bibinfo{person}{Zhiwei Guan} {and} \bibinfo{person}{Edward
  Cutrell}.} \bibinfo{year}{2007}\natexlab{}.
\newblock \showarticletitle{An eye tracking study of the effect of target rank
  on web search}. In \bibinfo{booktitle}{{\em Annual Conference of the ACM
  Special Interest Group on Computer Human Interaction}}. \bibinfo{address}{San
  Jose, California}.
\newblock


\bibitem[\protect\citeauthoryear{Guha, Cheng, and Francis}{Guha
  et~al\mbox{.}}{2010}]%
        {guha-2010-ads}
\bibfield{author}{\bibinfo{person}{Saikat Guha}, \bibinfo{person}{Bin Cheng},
  {and} \bibinfo{person}{Paul Francis}.} \bibinfo{year}{2010}\natexlab{}.
\newblock \showarticletitle{Challenges in Measuring Online Advertising
  Systems}. In \bibinfo{booktitle}{{\em ACM/USENIX Internet Measurement
  Conference}}. \bibinfo{address}{Melbourne, Victoria, Australia}.
\newblock


\bibitem[\protect\citeauthoryear{Hannak, Sapie\.{z}y\'{n}ski, Kakhki,
  Krishnamurthy, Lazer, Mislove, and Wilson}{Hannak et~al\mbox{.}}{2013}]%
        {hannak-2013-filterbubbles}
\bibfield{author}{\bibinfo{person}{Aniko Hannak}, \bibinfo{person}{Piotr
  Sapie\.{z}y\'{n}ski}, \bibinfo{person}{Arash~Molavi Kakhki},
  \bibinfo{person}{Balachander Krishnamurthy}, \bibinfo{person}{David Lazer},
  \bibinfo{person}{Alan Mislove}, {and} \bibinfo{person}{Christo Wilson}.}
  \bibinfo{year}{2013}\natexlab{}.
\newblock \showarticletitle{Measuring Personalization of Web Search}. In
  \bibinfo{booktitle}{{\em International World Wide Web Conference}}.
  \bibinfo{address}{Rio de Janeiro, Brazil}.
\newblock


\bibitem[\protect\citeauthoryear{Hines}{Hines}{2004}]%
        {hines-2004-personalsearch}
\bibfield{author}{\bibinfo{person}{Matt Hines}.}
  \bibinfo{year}{2004}\natexlab{}.
\newblock \bibinfo{title}{Google Takes Searching Personally}.
\newblock   (\bibinfo{date}{March} \bibinfo{year}{2004}).
\newblock
\newblock
\shownote{\url{http://cnet.co/V37pZD}.}


\bibitem[\protect\citeauthoryear{Horling and Kulick}{Horling and
  Kulick}{2009}]%
        {horling-2009-personalsearch}
\bibfield{author}{\bibinfo{person}{Bryan Horling} {and}
  \bibinfo{person}{Matthew Kulick}.} \bibinfo{year}{2009}\natexlab{}.
\newblock \bibinfo{title}{Personalized Search for Everyone}.
\newblock   (\bibinfo{date}{December} \bibinfo{year}{2009}).
\newblock
\newblock
\shownote{\url{http://bit.ly/71RcmJ}.}


\bibitem[\protect\citeauthoryear{Hu, Zeng, Li, Niu, and Chen}{Hu
  et~al\mbox{.}}{2007}]%
        {hu-2007-predicting}
\bibfield{author}{\bibinfo{person}{Jian Hu}, \bibinfo{person}{Hua-Jun Zeng},
  \bibinfo{person}{Hua Li}, \bibinfo{person}{Cheng Niu}, {and}
  \bibinfo{person}{Zheng Chen}.} \bibinfo{year}{2007}\natexlab{}.
\newblock \showarticletitle{Demographic Prediction Based on User's Browsing
  Behavior}. In \bibinfo{booktitle}{{\em International World Wide Web
  Conference}}. \bibinfo{address}{Banff, Canada}.
\newblock


\bibitem[\protect\citeauthoryear{Liu, Yu, and Meng}{Liu et~al\mbox{.}}{2002}]%
        {liu-2002-categories}
\bibfield{author}{\bibinfo{person}{Fang Liu}, \bibinfo{person}{Clement Yu},
  {and} \bibinfo{person}{Weiyi Meng}.} \bibinfo{year}{2002}\natexlab{}.
\newblock \showarticletitle{Personalized web search by mapping user queries to
  categories}. In \bibinfo{booktitle}{{\em ACM International Conference on
  Information and Knowledge Management}}. \bibinfo{address}{McLean, Virginia}.
\newblock


\bibitem[\protect\citeauthoryear{Liu, Yu, and Meng}{Liu et~al\mbox{.}}{2004}]%
        {liu-2004-websearch}
\bibfield{author}{\bibinfo{person}{Fang Liu}, \bibinfo{person}{Clement Yu},
  {and} \bibinfo{person}{Weiyi Meng}.} \bibinfo{year}{2004}\natexlab{}.
\newblock \showarticletitle{Personalized Web Search for Improving Retrieval
  Effectiveness}.
\newblock \bibinfo{journal}{{\em IEEE Transactions on Knowledge and Data
  Engineering\/}} \bibinfo{volume}{16}, \bibinfo{number}{1}
  (\bibinfo{date}{January} \bibinfo{year}{2004}), \bibinfo{pages}{28--40}.
\newblock


\bibitem[\protect\citeauthoryear{Lorigo, Pan, Hembrooke, Joachims, Granka, and
  Gay}{Lorigo et~al\mbox{.}}{2006}]%
        {lorigo-2006-influence}
\bibfield{author}{\bibinfo{person}{Lori Lorigo}, \bibinfo{person}{Bing Pan},
  \bibinfo{person}{Helene Hembrooke}, \bibinfo{person}{Thorsten Joachims},
  \bibinfo{person}{Laura Granka}, {and} \bibinfo{person}{Geri Gay}.}
  \bibinfo{year}{2006}\natexlab{}.
\newblock \showarticletitle{The influence of task and gender on search and
  evaluation behavior using Google}.
\newblock \bibinfo{journal}{{\em Information Processing and Management\/}}
  \bibinfo{volume}{42}, \bibinfo{number}{4} (\bibinfo{year}{2006}),
  \bibinfo{pages}{1123--1131}.
\newblock


\bibitem[\protect\citeauthoryear{Majumder and Shrivastava}{Majumder and
  Shrivastava}{2013}]%
        {majumder-2013-personalization}
\bibfield{author}{\bibinfo{person}{Anirban Majumder} {and}
  \bibinfo{person}{Nisheeth Shrivastava}.} \bibinfo{year}{2013}\natexlab{}.
\newblock \showarticletitle{Know your personalization: learning topic level
  personalization in online services}. In \bibinfo{booktitle}{{\em
  International World Wide Web Conference}}. \bibinfo{address}{Rio de Janeiro,
  Brazil}.
\newblock


\bibitem[\protect\citeauthoryear{Mattioli}{Mattioli}{2012}]%
        {mattioli-2012-orbitz}
\bibfield{author}{\bibinfo{person}{Dana Mattioli}.}
  \bibinfo{year}{2012}\natexlab{}.
\newblock \bibinfo{title}{On Orbitz, Mac Users Steered to Pricier Hotels}.
\newblock   (\bibinfo{date}{August} \bibinfo{year}{2012}).
\newblock
\newblock
\shownote{\url{http://on.wsj.com/LwTnPH}.}


\bibitem[\protect\citeauthoryear{Mikians, Gyarmati, Erramilli, and
  Laoutaris}{Mikians et~al\mbox{.}}{2012}]%
        {mikians-2012-prices}
\bibfield{author}{\bibinfo{person}{Jakub Mikians}, \bibinfo{person}{L\'aszl\'o
  Gyarmati}, \bibinfo{person}{Vijay Erramilli}, {and} \bibinfo{person}{Nikolaos
  Laoutaris}.} \bibinfo{year}{2012}\natexlab{}.
\newblock \showarticletitle{Detecting Price and Search Discrimination on the
  Internet}. In \bibinfo{booktitle}{{\em Workshop on Hot Topics in Networks}}.
  \bibinfo{address}{Seattle, Washington}.
\newblock


\bibitem[\protect\citeauthoryear{Mobasher, Cooley, and Srivastava}{Mobasher
  et~al\mbox{.}}{2000}]%
        {mobasher-2000-webusage}
\bibfield{author}{\bibinfo{person}{Bamshad Mobasher}, \bibinfo{person}{Robert
  Cooley}, {and} \bibinfo{person}{Jaideep Srivastava}.}
  \bibinfo{year}{2000}\natexlab{}.
\newblock \showarticletitle{Automatic Personalization Based on Web Usage
  Mining}.
\newblock \bibinfo{journal}{{\it Commun. ACM}} \bibinfo{volume}{43},
  \bibinfo{number}{8} (\bibinfo{year}{2000}), \bibinfo{pages}{142--151}.
\newblock


\bibitem[\protect\citeauthoryear{Pan, Hembrooke, Joachims, Lorigo, Gay, and
  Granka}{Pan et~al\mbox{.}}{2007}]%
        {pan-2007-trust}
\bibfield{author}{\bibinfo{person}{Bing Pan}, \bibinfo{person}{Helene
  Hembrooke}, \bibinfo{person}{Thorsten Joachims}, \bibinfo{person}{Lori
  Lorigo}, \bibinfo{person}{Geri Gay}, {and} \bibinfo{person}{Laura Granka}.}
  \bibinfo{year}{2007}\natexlab{}.
\newblock \showarticletitle{In Google We Trust: Users' Decisions on Rank,
  Position, and Relevance}.
\newblock \bibinfo{journal}{{\em Journal of Computer-Mediated Communication\/}}
  \bibinfo{volume}{12}, \bibinfo{number}{3} (\bibinfo{year}{2007}),
  \bibinfo{pages}{801--823}.
\newblock


\bibitem[\protect\citeauthoryear{Panopticlick}{Panopticlick}{}]%
        {Panopticlick}
Panopticlick.
\newblock \bibinfo{title}{Panopticlick}.
\newblock   (\bibinfo{year}{????}).
\newblock
\newblock
\shownote{\url{https://panopticlick.eff.org}.}


\bibitem[\protect\citeauthoryear{Pansari and Mayer}{Pansari and Mayer}{2006}]%
        {pansari-2006-abtest}
\bibfield{author}{\bibinfo{person}{Ambar Pansari} {and}
  \bibinfo{person}{Marissa Mayer}.} \bibinfo{year}{2006}\natexlab{}.
\newblock \bibinfo{title}{This is a test. This is only a test.}
\newblock   (\bibinfo{date}{April} \bibinfo{year}{2006}).
\newblock
\newblock
\shownote{\url{http://bit.ly/Ldbb0}.}


\bibitem[\protect\citeauthoryear{Pariser}{Pariser}{2011}]%
        {pariser-2011-filterbubble}
\bibfield{author}{\bibinfo{person}{Eli Pariser}.}
  \bibinfo{year}{2011}\natexlab{}.
\newblock \bibinfo{booktitle}{{\em The Filter Bubble: What the Internet is
  Hiding from You}}.
\newblock \bibinfo{publisher}{Penguin Press}.
\newblock


\bibitem[\protect\citeauthoryear{PhantomJS}{PhantomJS}{2013}]%
        {PhantomJS}
PhantomJS \bibinfo{year}{2013}\natexlab{}.
\newblock \bibinfo{title}{PhantomJS}.
\newblock   (\bibinfo{year}{2013}).
\newblock
\newblock
\shownote{\url{http://phantomjs.org}.}


\bibitem[\protect\citeauthoryear{Pitkow, Sch\"{u}tze, Cass, Cooley, Turnbull,
  Edmonds, Adar, and Breuel}{Pitkow et~al\mbox{.}}{2002}]%
        {pitkow-2002-personalized}
\bibfield{author}{\bibinfo{person}{James Pitkow}, \bibinfo{person}{Hinrich
  Sch\"{u}tze}, \bibinfo{person}{Todd Cass}, \bibinfo{person}{Rob Cooley},
  \bibinfo{person}{Don Turnbull}, \bibinfo{person}{Andy Edmonds},
  \bibinfo{person}{Eytan Adar}, {and} \bibinfo{person}{Thomas Breuel}.}
  \bibinfo{year}{2002}\natexlab{}.
\newblock \showarticletitle{Personalized search}.
\newblock \bibinfo{journal}{{\it Commun. ACM}} \bibinfo{volume}{45},
  \bibinfo{number}{9} (\bibinfo{date}{September} \bibinfo{year}{2002}),
  \bibinfo{pages}{50--55}.
\newblock


\bibitem[\protect\citeauthoryear{Pretschner and Gauch}{Pretschner and
  Gauch}{1999}]%
        {pretschner-1999-personalize}
\bibfield{author}{\bibinfo{person}{Alexander Pretschner} {and}
  \bibinfo{person}{Susan Gauch}.} \bibinfo{year}{1999}\natexlab{}.
\newblock \showarticletitle{Ontology based personalized search}. In
  \bibinfo{booktitle}{{\em IEEE International Conference on Tools with
  Artificial Intelligence}}. \bibinfo{address}{Chicago, Illinois}.
\newblock


\bibitem[\protect\citeauthoryear{Qiu and Cho}{Qiu and Cho}{2006}]%
        {qiu-2006-personalized}
\bibfield{author}{\bibinfo{person}{Feng Qiu} {and} \bibinfo{person}{Junghoo
  Cho}.} \bibinfo{year}{2006}\natexlab{}.
\newblock \showarticletitle{Automatic Identification of User Interest for
  Personalized Search}. In \bibinfo{booktitle}{{\em International World Wide
  Web Conference}}. \bibinfo{address}{Edinburgh, Scotland}.
\newblock


\bibitem[\protect\citeauthoryear{Quantcast}{Quantcast}{2012}]%
        {quantcast-2012}
\bibfield{author}{\bibinfo{person}{Quantcast}.}
  \bibinfo{year}{2012}\natexlab{}.
\newblock \bibinfo{title}{Top Sites for the United States}.
\newblock   (\bibinfo{year}{2012}).
\newblock
\newblock
\shownote{\url{http://www.quantcast.com/top-sites}.}


\bibitem[\protect\citeauthoryear{Robinson}{Robinson}{2013}]%
        {robinson-2013-googvseu}
\bibfield{author}{\bibinfo{person}{Frances Robinson}.}
  \bibinfo{year}{2013}\natexlab{}.
\newblock \bibinfo{title}{EU Tells Google to Offer More in Search Probe}.
\newblock   (\bibinfo{date}{July} \bibinfo{year}{2013}).
\newblock
\newblock
\shownote{\url{http://online.wsj.com/article/SB10001424127887323993804578611362017999002.html}.}


\bibitem[\protect\citeauthoryear{Roesner, Kohno, and Wetherall}{Roesner
  et~al\mbox{.}}{2012}]%
        {roesner-2012-tracking}
\bibfield{author}{\bibinfo{person}{Franziska Roesner},
  \bibinfo{person}{Tadayoshi Kohno}, {and} \bibinfo{person}{David Wetherall}.}
  \bibinfo{year}{2012}\natexlab{}.
\newblock \showarticletitle{Detecting and Defending Against Third-Party
  Tracking on the Web}. In \bibinfo{booktitle}{{\em Symposium on Networked
  System Design and Implementation}}. \bibinfo{address}{San Jose, California}.
\newblock


\bibitem[\protect\citeauthoryear{Selenium}{Selenium}{2013}]%
        {Selenium}
Selenium \bibinfo{year}{2013}\natexlab{}.
\newblock \bibinfo{title}{Selenium}.
\newblock   (\bibinfo{year}{2013}).
\newblock
\newblock
\shownote{\url{http://selenium.org}.}


\bibitem[\protect\citeauthoryear{Shen, Tan, and Zhai}{Shen
  et~al\mbox{.}}{2005}]%
        {shen-2005-modeling}
\bibfield{author}{\bibinfo{person}{Xuehua Shen}, \bibinfo{person}{Bin Tan},
  {and} \bibinfo{person}{ChengXiang Zhai}.} \bibinfo{year}{2005}\natexlab{}.
\newblock \showarticletitle{Implicit user modeling for personalized search}. In
  \bibinfo{booktitle}{{\em ACM International Conference on Information and
  Knowledge Management}}. \bibinfo{address}{Bremen, Germany}.
\newblock


\bibitem[\protect\citeauthoryear{Shen, Yan, Yan, Ji, Liu, and Chen}{Shen
  et~al\mbox{.}}{2011}]%
        {shen-2011-intent}
\bibfield{author}{\bibinfo{person}{Yelong Shen}, \bibinfo{person}{Jun Yan},
  \bibinfo{person}{Shuicheng Yan}, \bibinfo{person}{Lei Ji},
  \bibinfo{person}{Ning Liu}, {and} \bibinfo{person}{Zheng Chen}.}
  \bibinfo{year}{2011}\natexlab{}.
\newblock \showarticletitle{Sparse Hidden-Dynamics Conditional Random Fields
  for User Intent Understanding}. In \bibinfo{booktitle}{{\em International
  World Wide Web Conference}}. \bibinfo{address}{Hyderabad, India}.
\newblock


\bibitem[\protect\citeauthoryear{Sieg, Mobasher, and Burke}{Sieg
  et~al\mbox{.}}{2007}]%
        {sieg-2007-ontological}
\bibfield{author}{\bibinfo{person}{Ahu Sieg}, \bibinfo{person}{Bamshad
  Mobasher}, {and} \bibinfo{person}{Robin Burke}.}
  \bibinfo{year}{2007}\natexlab{}.
\newblock \showarticletitle{Web Search Personalization with Ontological User
  Profiles}. In \bibinfo{booktitle}{{\em ACM International Conference on
  Information and Knowledge Management}}.
\newblock


\bibitem[\protect\citeauthoryear{Singer}{Singer}{2011}]%
        {singer-2011-filterbubble}
\bibfield{author}{\bibinfo{person}{Natasha Singer}.}
  \bibinfo{year}{2011}\natexlab{}.
\newblock \bibinfo{title}{The Trouble with the Echo Chamber Online}.
\newblock   (\bibinfo{date}{May} \bibinfo{year}{2011}).
\newblock
\newblock
\shownote{\url{http://nyti.ms/jcTih2}.}


\bibitem[\protect\citeauthoryear{Singhal}{Singhal}{2011}]%
        {singhal-2011-personalsearch}
\bibfield{author}{\bibinfo{person}{Amit Singhal}.}
  \bibinfo{year}{2011}\natexlab{}.
\newblock \bibinfo{title}{Some Thoughts on Personalization}.
\newblock   (\bibinfo{date}{November} \bibinfo{year}{2011}).
\newblock
\newblock
\shownote{\url{http://bit.ly/tJS4xT}.}


\bibitem[\protect\citeauthoryear{Singhal}{Singhal}{2012}]%
        {singhal-2012-personalsearch}
\bibfield{author}{\bibinfo{person}{Amit Singhal}.}
  \bibinfo{year}{2012}\natexlab{}.
\newblock \bibinfo{title}{Search, Plus Your World}.
\newblock   (\bibinfo{date}{January} \bibinfo{year}{2012}).
\newblock
\newblock
\shownote{\url{http://bit.ly/yUJnCl}.}


\bibitem[\protect\citeauthoryear{Slegg}{Slegg}{2013}]%
        {search_share-2013}
\bibfield{author}{\bibinfo{person}{Jennifer Slegg}.}
  \bibinfo{year}{2013}\natexlab{}.
\newblock \bibinfo{title}{Google, Bing Both Win More Search Market Share}.
\newblock   (\bibinfo{date}{June} \bibinfo{year}{2013}).
\newblock
\newblock
\shownote{\url{http://bit.ly/1bSllOR}.}


\bibitem[\protect\citeauthoryear{Smith, Linden, and Zada}{Smith
  et~al\mbox{.}}{2005}]%
        {smith-2005-content}
\bibfield{author}{\bibinfo{person}{B.~R. Smith}, \bibinfo{person}{G.~D.
  Linden}, {and} \bibinfo{person}{N.~K. Zada}.}
  \bibinfo{year}{2005}\natexlab{}.
\newblock \bibinfo{title}{Content personalization based on actions performed
  during a current browsing session}.
\newblock   (\bibinfo{date}{February} \bibinfo{year}{2005}).
\newblock
\showURL{%
\url{http://www.google.com/patents/US6853982}}
\newblock
\shownote{US Patent 6,853,982.}


\bibitem[\protect\citeauthoryear{Sullivan}{Sullivan}{2011}]%
        {bing_personal-2011}
\bibfield{author}{\bibinfo{person}{Danny Sullivan}.}
  \bibinfo{year}{2011}\natexlab{}.
\newblock \bibinfo{title}{Bing Results Get Local and Personalized}.
\newblock   (\bibinfo{date}{February} \bibinfo{year}{2011}).
\newblock
\newblock
\shownote{\url{http://selnd.com/hY4djp}.}


\bibitem[\protect\citeauthoryear{Sullivan}{Sullivan}{2012}]%
        {sullivan-2012-personalization}
\bibfield{author}{\bibinfo{person}{Danny Sullivan}.}
  \bibinfo{year}{2012}\natexlab{}.
\newblock \bibinfo{title}{Why Google ``Personalizes'' Results Based on Obama
  Searches But Not Romney}.
\newblock   (\bibinfo{date}{November} \bibinfo{year}{2012}).
\newblock
\newblock
\shownote{\url{http://selnd.com/PyfvvY}.}


\bibitem[\protect\citeauthoryear{Sun, Zeng, Liu, Lu, and Chen}{Sun
  et~al\mbox{.}}{2005}]%
        {sun-2005-personalized}
\bibfield{author}{\bibinfo{person}{Jian-Tao Sun}, \bibinfo{person}{Hua-Jun
  Zeng}, \bibinfo{person}{Huan Liu}, \bibinfo{person}{Yuchang Lu}, {and}
  \bibinfo{person}{Zheng Chen}.} \bibinfo{year}{2005}\natexlab{}.
\newblock \showarticletitle{CubeSVD: A Novel Approach to Personalized Web
  Search}. In \bibinfo{booktitle}{{\em International World Wide Web
  Conference}}. \bibinfo{address}{Chiba, Japan}.
\newblock


\bibitem[\protect\citeauthoryear{Sun, Lebanon, and Collins-Thompson}{Sun
  et~al\mbox{.}}{2010}]%
        {sun-2010-visualizing}
\bibfield{author}{\bibinfo{person}{Mingxuan Sun}, \bibinfo{person}{Guy
  Lebanon}, {and} \bibinfo{person}{Kevyn Collins-Thompson}.}
  \bibinfo{year}{2010}\natexlab{}.
\newblock \showarticletitle{Visualizing Differences in Web Search Algorithms
  using the Expected Weighted Hoeffding Distance}. In \bibinfo{booktitle}{{\em
  International World Wide Web Conference}}. \bibinfo{address}{Raleigh, North
  Carolina}.
\newblock


\bibitem[\protect\citeauthoryear{Sydell}{Sydell}{2012}]%
        {RickSantorumGoogle}
\bibfield{author}{\bibinfo{person}{Laura Sydell}.}
  \bibinfo{year}{2012}\natexlab{}.
\newblock \bibinfo{title}{How Rick Santorum's `Google Problem' Has Endured}.
\newblock   (\bibinfo{year}{2012}).
\newblock
\newblock
\shownote{\url{http://n.pr/wefdnc}.}


\bibitem[\protect\citeauthoryear{Tan, Shen, and Zhai}{Tan
  et~al\mbox{.}}{2006}]%
        {tan-2006-personalized}
\bibfield{author}{\bibinfo{person}{Bin Tan}, \bibinfo{person}{Xuehua Shen},
  {and} \bibinfo{person}{ChengXiang Zhai}.} \bibinfo{year}{2006}\natexlab{}.
\newblock \showarticletitle{Mining long-term search history to improve search
  accuracy}. In \bibinfo{booktitle}{{\em ACM SIGKDD International Conference of
  Knowledge Discovery and Data Mining}}. \bibinfo{address}{Philadelphia,
  Pennsylvania}.
\newblock


\bibitem[\protect\citeauthoryear{Teevan, Dumais, and Horvitz}{Teevan
  et~al\mbox{.}}{2005}]%
        {teevan-2005-personalized}
\bibfield{author}{\bibinfo{person}{Jamie Teevan}, \bibinfo{person}{Susan~T.
  Dumais}, {and} \bibinfo{person}{Eric Horvitz}.}
  \bibinfo{year}{2005}\natexlab{}.
\newblock \showarticletitle{Personalizing search via automated analysis of
  interests and activities}. In \bibinfo{booktitle}{{\em Conference of the ACM
  Special Interest Group on Information Retrieval}}.
  \bibinfo{address}{Sheffield, United Kingdom}.
\newblock


\bibitem[\protect\citeauthoryear{Vaughan}{Vaughan}{2004}]%
        {vaughan-2004-measurements}
\bibfield{author}{\bibinfo{person}{Liwen Vaughan}.}
  \bibinfo{year}{2004}\natexlab{}.
\newblock \showarticletitle{New measurements for search engine evaluation
  proposed and tested}.
\newblock \bibinfo{journal}{{\em Information Processing and Management\/}}
  \bibinfo{volume}{40}, \bibinfo{number}{4} (\bibinfo{date}{May}
  \bibinfo{year}{2004}), \bibinfo{pages}{677--691}.
\newblock


\bibitem[\protect\citeauthoryear{WebMD Year in Health}{WebMD Year in
  Health}{2011}]%
        {WebMDSearches}
WebMD Year in Health \bibinfo{year}{2011}\natexlab{}.
\newblock \bibinfo{title}{WebMD Year in Health}.
\newblock   (\bibinfo{year}{2011}).
\newblock
\newblock
\shownote{\url{http://on.webmd.com/eBPFxH}.}


\bibitem[\protect\citeauthoryear{White}{White}{2013}]%
        {white-2013-websearchbias}
\bibfield{author}{\bibinfo{person}{Ryen~W. White}.}
  \bibinfo{year}{2013}\natexlab{}.
\newblock \showarticletitle{Beliefs and Biases in Web Search}. In
  \bibinfo{booktitle}{{\em Conference of the ACM Special Interest Group on
  Information Retrieval}}.
\newblock


\bibitem[\protect\citeauthoryear{Wills and Tatar}{Wills and Tatar}{2012}]%
        {wills-2012-ads}
\bibfield{author}{\bibinfo{person}{Craig~E. Wills} {and} \bibinfo{person}{Can
  Tatar}.} \bibinfo{year}{2012}\natexlab{}.
\newblock \showarticletitle{Understanding What They Do with What They Know}. In
  \bibinfo{booktitle}{{\em Workshop on Privacy in the Electronic Society}}.
\newblock


\bibitem[\protect\citeauthoryear{Wines}{Wines}{2012}]%
        {GoogleChina}
\bibfield{author}{\bibinfo{person}{Michael Wines}.}
  \bibinfo{year}{2012}\natexlab{}.
\newblock \bibinfo{title}{Google to Alert Users to Chinese Censorship}.
\newblock   (\bibinfo{date}{June} \bibinfo{year}{2012}).
\newblock
\newblock
\shownote{{http://nyti.ms/JRhGZS}.}


\bibitem[\protect\citeauthoryear{Witten}{Witten}{2012}]%
        {whitten-2012-privacypolicy}
\bibfield{author}{\bibinfo{person}{Alma Witten}.}
  \bibinfo{year}{2012}\natexlab{}.
\newblock \bibinfo{title}{Google's New Privacy Policy}.
\newblock   (\bibinfo{date}{March} \bibinfo{year}{2012}).
\newblock
\newblock
\shownote{\url{http://bit.ly/wVr4mF}.}


\bibitem[\protect\citeauthoryear{Xu, Zhang, Chen, and Wang}{Xu
  et~al\mbox{.}}{2007}]%
        {xu-2007-private}
\bibfield{author}{\bibinfo{person}{Yabo Xu}, \bibinfo{person}{Benyu Zhang},
  \bibinfo{person}{Zheng Chen}, {and} \bibinfo{person}{Ke Wang}.}
  \bibinfo{year}{2007}\natexlab{}.
\newblock \showarticletitle{Privacy-Enhancing Personalized Web Search}. In
  \bibinfo{booktitle}{{\em International World Wide Web Conference}}.
  \bibinfo{address}{Banff, Canada}.
\newblock


\bibitem[\protect\citeauthoryear{Yi, Raghavan, and Leggetter}{Yi
  et~al\mbox{.}}{2009}]%
        {yi-2009-geointention}
\bibfield{author}{\bibinfo{person}{Xing Yi}, \bibinfo{person}{Hema Raghavan},
  {and} \bibinfo{person}{Chris Leggetter}.} \bibinfo{year}{2009}\natexlab{}.
\newblock \showarticletitle{Discovering Users' Specific Geo Intention in Web
  Search}. In \bibinfo{booktitle}{{\em International World Wide Web
  Conference}}. \bibinfo{address}{Madrid, Spain}.
\newblock


\bibitem[\protect\citeauthoryear{Yu and Cai}{Yu and Cai}{2007}]%
        {yu-2007-geographic}
\bibfield{author}{\bibinfo{person}{Bo Yu} {and} \bibinfo{person}{Guoray Cai}.}
  \bibinfo{year}{2007}\natexlab{}.
\newblock \showarticletitle{A query-aware document ranking method for
  geographic information retrieval}. In \bibinfo{booktitle}{{\em ACM Workshop
  On Geographic Information Retrieval}}. \bibinfo{address}{Lisbon, Portugal}.
\newblock


\end{thebibliography}

\end{document}